 \documentclass[preprint,12pt]{elsarticle}
\usepackage[top=2.54cm, bottom=2.54cm, left=2.54cm, right=2.54cm]{geometry}

\usepackage{amssymb}
\usepackage{amsmath}

\usepackage{comment}
\usepackage{graphicx}
\usepackage{multirow}
\usepackage{booktabs}
\usepackage{parskip}
\usepackage{rotating}
\usepackage{dcolumn}
\usepackage{float}
\usepackage[nottoc,numbib]{tocbibind}
\usepackage{textcomp}
\usepackage{subfigure}
\usepackage{xspace}
\usepackage{tabularx}
\usepackage{adjustbox}
\usepackage[all, defaultlines=4]{nowidow}
\usepackage{wrapfig}
\usepackage{afterpage}
\usepackage{array}
\usepackage{placeins}
\usepackage{hyperref}
\usepackage{siunitx}

\begin{document}

\begin{frontmatter}


\title{Measurement of the cosmic muon flux at the Stawell Underground Physics Laboratory}


\author[1,2]{G.~Fu}
\author[1,2]{M.~Mews}
\author[1,6]{F.~Scutti \corref{cor1}}
\ead{fscutti@swin.edu.au}
\author[1,2]{P.~Urquijo}

\author[1,2]{E.~Barberio}
\author[1,3]{V.~Bashu}
\author[1,3]{L.J.~Bignell}
\author[1,4,12]{I.~Bolognino}
\author[1,2]{A.~Cools}
\author[1,3]{F.~Dastgiri}
\author[1,6,11]{A.R.~Duffy}
\author[1,2]{L.~Einfalt}
\author[1,3]{M.~Froehlich}
\author[1,8]{T.~Fruth}
\author[1,2]{M.~Gerathy}
\author[1,4]{M.~Hancock}
\author[1,2]{R.~James}
\author[1,8]{S.~Kapoor}
\author[1,6]{S.~Krishnan}
\author[1,3]{G.J.~Lane}
\author[1,4]{K.T.~Leaver}
\author[1,2]{D.~Marcantonio}
\author[1,4]{P.~McGee}
\author[1,2]{J.~McKenzie}
\author[1,3]{L.~McKie}
\author[9,10]{M.A.~McLean}
\author[1,3]{P.C.~McNamara}
\author[1,2]{L.J.~Milligan}
\author[1,2]{K.J.~Rule}
\author[1,3]{Z.~Slavkovsk\'{a}}
\author[1,2]{O.~Stanley}
\author[1,3]{A.E.~Stuchbery}
\author[1,7]{B.~Suerfu}
\author[1,2]{G.N.~Taylor}
\author[1,11]{E.~{van der}~Velden}
\author[1,4]{A.G.~Williams}
\author[1,2]{Y.~Xing}
\author[1,3]{Y.Y.~Zhong}

\cortext[cor1]{Corresponding author }

\affiliation[1]{organization={ARC Centre of Excellence for Dark Matter Particle Physics},
            country={Australia}}
            
\affiliation[2]{organization={School of Physics, The University of Melbourne},
            addressline={Parkville}, 
            city={Melbourne},
            postcode={3010}, 
            state={VIC},
            country={Australia}}

\affiliation[3]{organization={Department of Nuclear Physics and Accelerator Applications, The Australian National University},
             city={Canberra},
             postcode={2601},
             state={ACT},
             country={Australia}}

\affiliation[4]{organization={Department of Physics, The University of Adelaide},
             city={Adelaide},
             postcode={5005},
             state={SA},
             country={Australia}}

\affiliation[6]{organization={Centre for Astrophysics and Supercomputing, Swinburne University of Technology},
             addressline={Hawthorn},   
             city={Melbourne},
             postcode={3122},
             state={VIC},
             country={Australia}}

\affiliation[7]{organization={High Energy Accelerator Research Organization},
             addressline={Oho},
             city={Tsukuba},
             postcode={305-0801},
             state={Ibaraki},
             country={Japan}}

\affiliation[8]{organization={School of Physics, The University of Sydney},
             addressline={Camperdown},
             city={Sydney},
             postcode={2006},
             state={NSW},
             country={Australia}}
             
\affiliation[9]{organization={School of Geography, Earth and Atmospheric Sciences, The University of Melbourne},
             addressline={Parkville},
             city={Melbourne},
             postcode={3010},
             state={VIC},
             country={Australia}}

\affiliation[10]{organization={Geological Survey of Victoria},
             addressline={East Melbourne},
             city={Melbourne},
             postcode={3002},
             state={VIC},
             country={Australia}}

\affiliation[11]{organization={ARC Centre of Excellence for All Sky Astrophysics in 3 Dimensions (ASTRO 3D)},
             city={Canberra},
             postcode={2611},
             state={ACT},
             country={Australia}}

\affiliation[12]{organization={INFN Sezione di Milano},
             addressline={via Celoria 16},
             city={Milano},
             postcode={20133},
             country={Italy}}

\begin{abstract}
We report the first measurement of the underground cosmic muon flux at the Stawell Underground Physics Laboratory. The measurement uses eight EJ200 plastic scintillator panels, equipped with Hamamatsu R13089 PMT pairs at the ends, which are the primary components of the muon veto system for the upcoming SABRE South experiment.  
The measured muon flux is $f = (6.33\,\pm\,0.04_{\rm stat}\,\pm\,0.35_{\rm sys})\times10^{-8}\,[{\rm s^{-1} cm^{-2}}].$ This measurement is in excellent agreement with simulations, with a relative uncertainty an order of magnitude smaller than the modelling uncertainty.
\end{abstract}


\end{frontmatter}

\section{Introduction}
\label{sec:introduction}

The Stawell Underground Physics Laboratory (SUPL) is a new underground research facility situated in Stawell, Victoria, Australia, and the first deep underground physics laboratory in the Southern Hemisphere. In the coming decade, it will host several particle physics experiments dedicated to detecting rare events, such as the upcoming SABRE South experiment~\cite{SABRESouth:2024bpv} aiming at dark matter direct detection. This work describes the first physics milestone of the laboratory: the measurement of the cosmic muon flux. This baseline quantity is a crucial component of the environmental background affecting all the experiments that will be hosted at SUPL. For the SABRE South experiment, it additionally represents an in-situ test of its data acquisition system and data processing pipeline. 

SUPL is located 1025\,m below ground, featuring a flat overburden. The laboratory is situated as an offshoot of the operationally active Stawell Gold Mine, at coordinates $37^\circ03' 38''\mathrm{S}$, $142^\circ 47' 59''\mathrm{E}$ with a surface elevation of approximately 275\,m above sea level. The low cosmic ray and radioactivity environment offered by the laboratory is an essential requirement for the sensitive experiments it will host. However, natural radiation sources that produce neutrons and gamma rays will need to be well understood. This work measures the cosmic muon flux at SUPL. This measurement is performed using the muon veto system of the SABRE South experiment, arranged in the telescopic configuration, using approximately one year of data collected between 2024 and 2025.

\section{Detector setup and data acquisition}
\label{sec:setup_and_daq}

\subsection{Muon detector setup}
\label{subsec:muon_detector_setup}
\begin{figure}[htbp]
    \centering
    \includegraphics[width=1.0\textwidth]{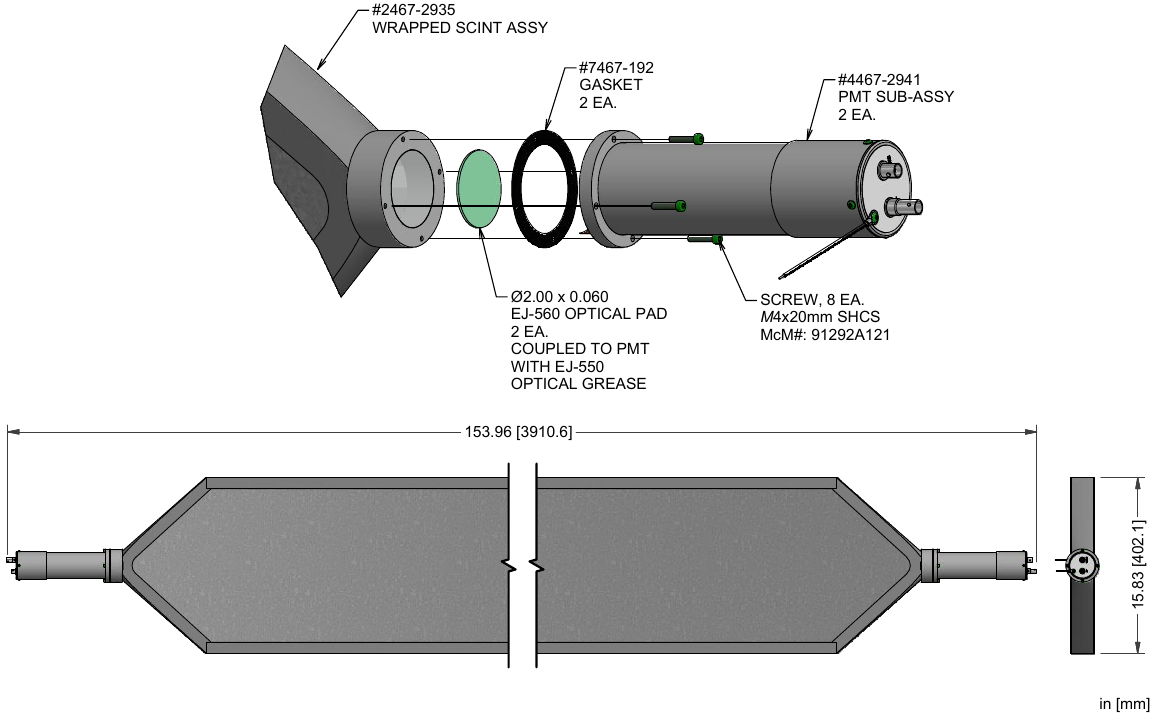}
    \caption{Assembly of a muon detector panel from the SABRE South muon detector veto system. The picture is provided by Eljen Technology~\cite{Eljen}.}
    \label{fig:muonpanel-assembly}
\end{figure}

\begin{figure}[htbp]
    \centering
    \includegraphics[width=0.4\textwidth]{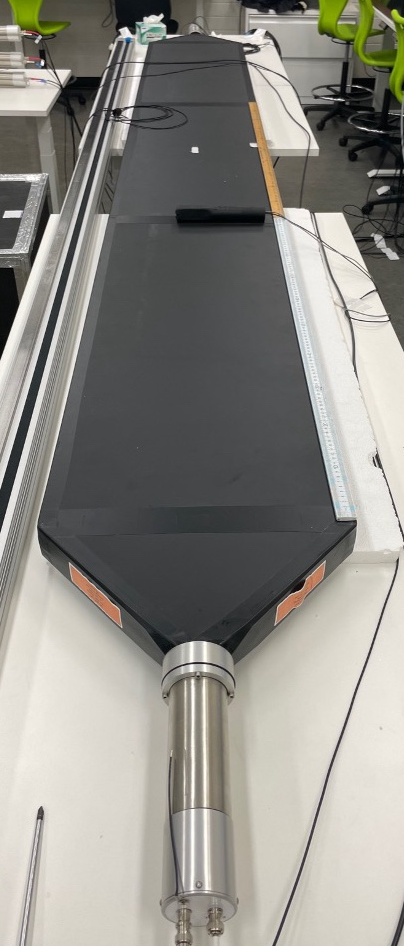}
    \caption{Muon scintillator panel under test in the above-ground laboratory at the University of Melbourne.}
    \label{fig:muonpanel-photo}
\end{figure}

The muon veto system of the SABRE South experiment consists of eight 5\,cm-thick Eljen EJ-200 plastic scintillator panels. Each panel has a sensitive area measuring 3000\,mm in length, 400\,mm in width, and 50\,mm in depth. The panels will be arranged in a flat array covering 9.6 m$^2$ above the active volume of the experiment to tag and veto muon events. Each muon detector panel is coupled to two 2-inch round Hamamatsu R13089 photomultiplier tubes (PMTs), with one positioned at each end. This configuration provides optimal detection efficiency and enables muon position reconstruction using timing information from the two PMTs. The schematic of a muon detector panel is shown in Fig.~\ref{fig:muonpanel-assembly} together with a PMT assembly. Each panel features diamond-milled edges, cast and polished faces, and an acrylic fishtail-shaped light guide, secured with EJ-500 optical cement. PMT adapters are glued to the other end of each light guide with WELDON-40 acrylic cement and fastened to the housing via flanges. The cavity between the PMT mount flanges and the adapters is filled with black Room Temperature Vulcanising (RTV) silicone sealant to ensure a proper seal. The PMTs are mounted to the adapters via the flanges, with EJ-560 optical pads and EJ-550 optical grease serving as the interface. A gasket between the two flanges ensures effective sealing and mechanical cushioning. To maximise light collection efficiency, each panel is wrapped in aluminium reflective foil. It is also wrapped with black vinyl to protect and prevent light leakage. The modules were constructed by Eljen Technology and shipped to Melbourne, where they were tested in a ground-level laboratory. A panel coupled to two PMTs is shown in Fig.~\ref{fig:muonpanel-photo}. Data collected above ground is used to measure the single-panel muon detection efficiency as described in Sec.~\ref{subsec:detection_efficiency}. 

\begin{figure}[htbp]
    \centering
\includegraphics[width=0.42\textwidth]{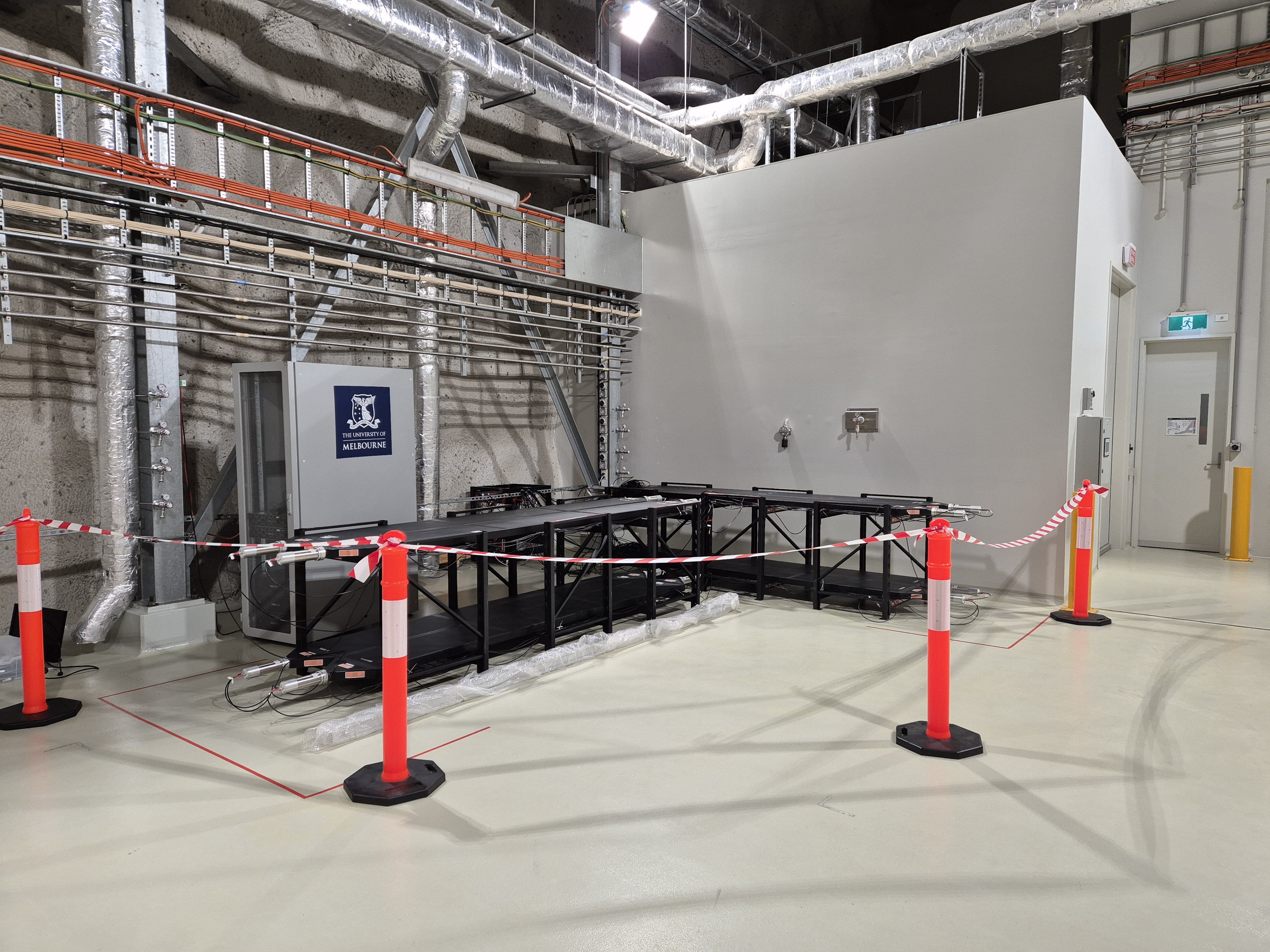}
\includegraphics[width=0.45\textwidth]{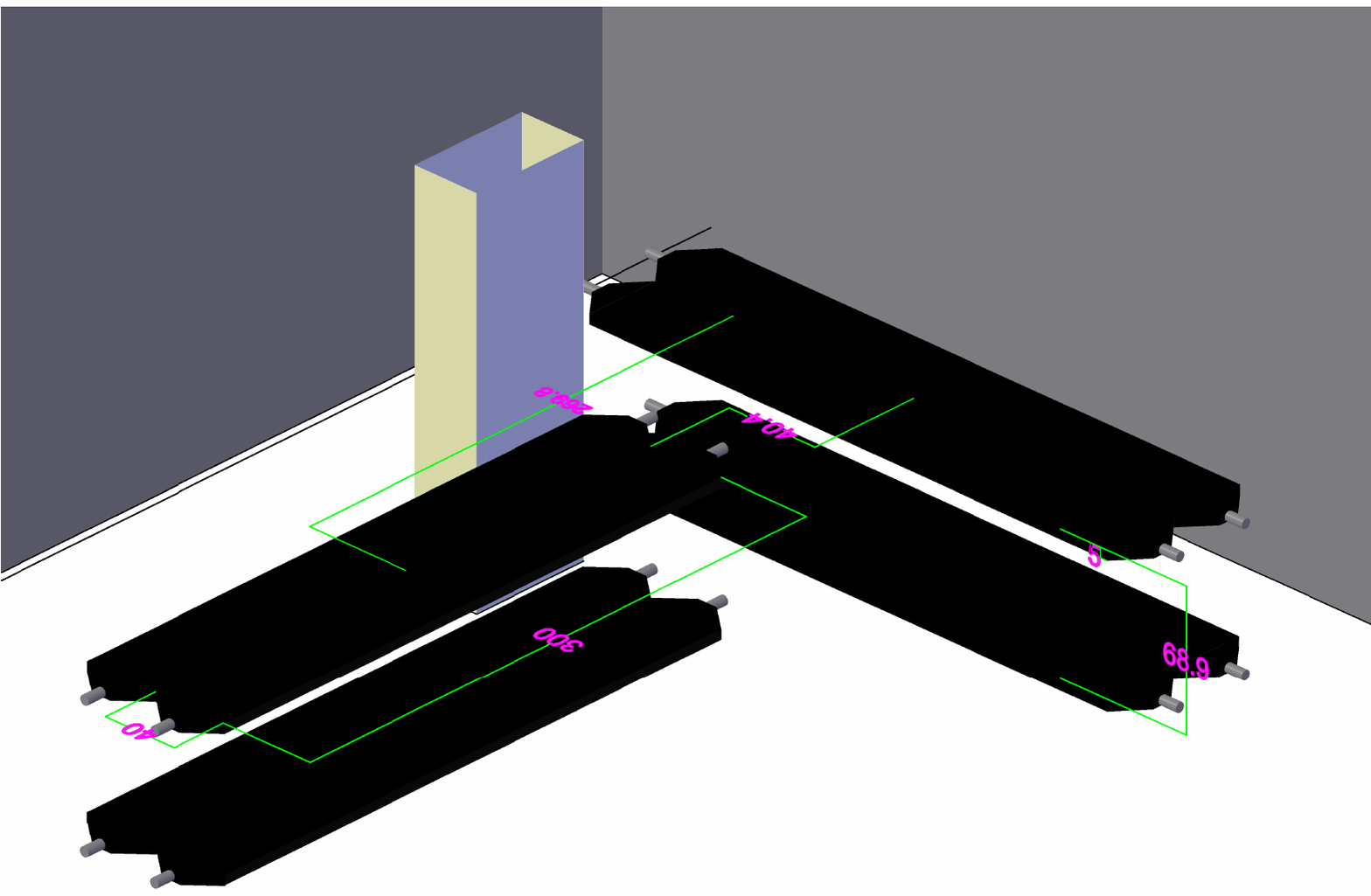}
\caption{\label{fig:config_SUPL}\raggedright (a)~The muon detector configuration at SUPL for the muon flux measurements. Eight muon detector panels are deployed in two layers and two orientations. (b)~A diagram of the muon telescope configuration for the muon flux measurement. }
\end{figure}

The SABRE South muon veto system will feature scintillator panels arranged in the same plane. However, for the present study, these are placed in a telescope configuration, defined as an arrangement allowing the measurement of the incident muon angle. Two telescopes are configured with orthogonal orientations to each other. Each telescope comprises two layers 68.9\,cm apart, each featuring pairs of scintillator panels positioned adjacent to each other. This setup is shown in Fig.~\ref{fig:config_SUPL}. While angular-dependent flux measurements are beyond the scope of this study, they will be documented in future studies from the SABRE South collaboration. 

All 16 Hamamatsu R13089 PMTs are connected to a 16-channel CAEN V1743 digitiser with a dynamic range of 2.5\,V and a resolution of 12 bits. Each waveform is sampled at a rate of 3.2\,GS/s, with a total of 1024 samples recorded over a time interval of 320\,ns. Each event is followed by a 125\,$\mu$s dead time period, as specified by the manufacturer, due to the analog-to-digital conversion process implemented by the digitiser. For each channel, the trigger threshold is set to 10\,ADC, corresponding to 6.1\,mV. All PMTs are powered by a CAEN SY5527 mainframe equipped with an A7435SN high-voltage (HV) board. 

Single-panel events are triggered by coincident signals between two channels corresponding to the same panel. The gate width for this trigger is 50\,ns. However, multi-panel events require triggers between at least two single-panel events. The multi-panel trigger window is set to 120\,ns. When multi-panel triggers are issued, the waveforms from all 16 channels are recorded for further analysis.

\subsection{Environmental monitoring}
\label{subsec:environmental_monitoring}
\begin{figure}[htb]
    \centering
    \includegraphics[width=0.6\linewidth]{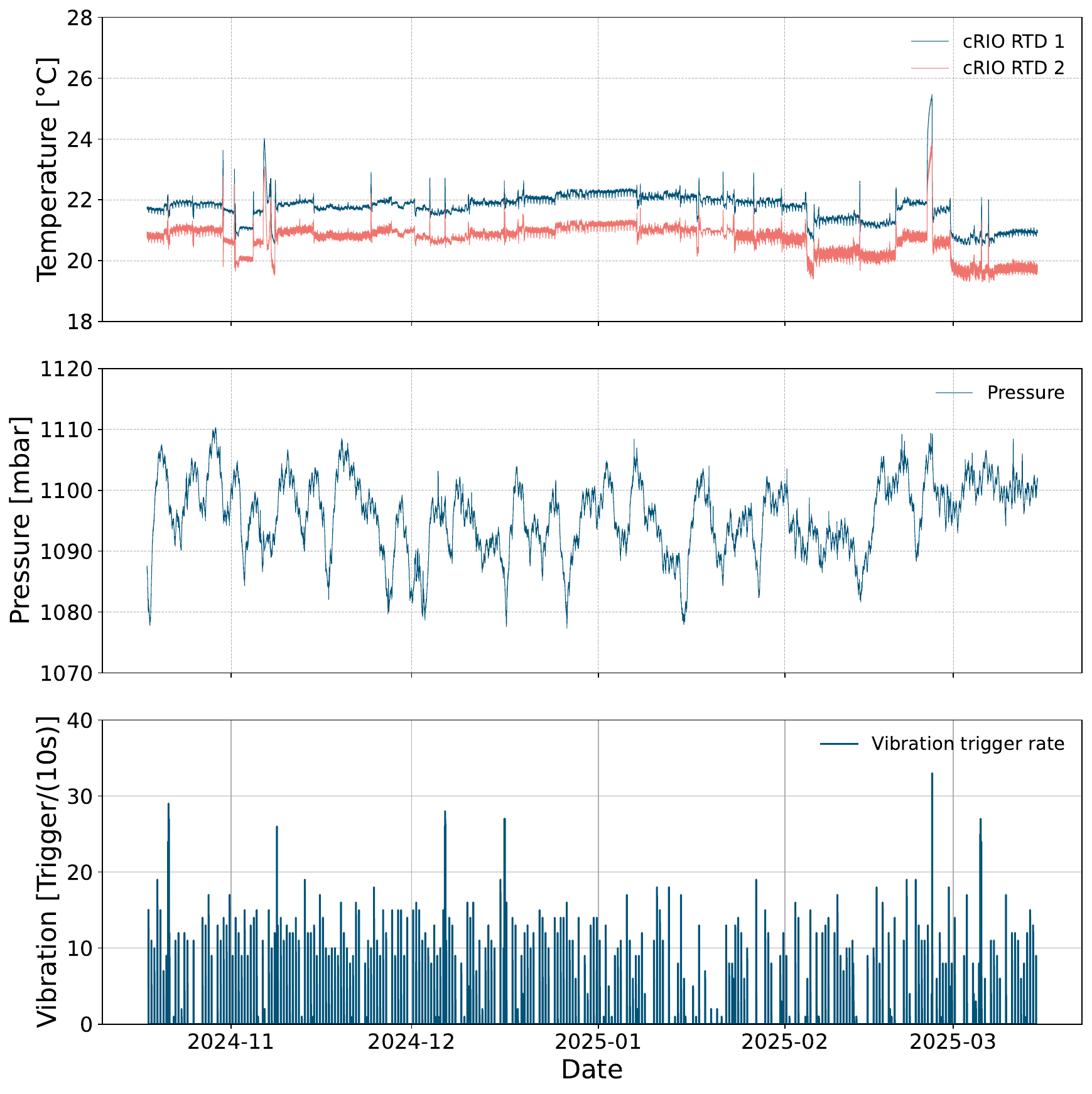}
    \caption{Temperature (top), pressure (middle), and vibration trigger rate (bottom) data monitored by the SABRE South slow control system.}
    \label{fig:SC_TempPress}
\end{figure}

As part of the SABRE South experiment, a prototype slow control system has been developed as an integral part of the data quality assessment~\cite{Krishnan2021}. The prototype slow control system has been used for the measurement presented here. 
The system monitors various experimental conditions, including temperature, pressure, and seismic vibration. 
Figure~\ref{fig:SC_TempPress} shows an example of temperature, pressure, and vibration data readings from October 2024 to March 2025, during which the slow control system was operational. 

\subsection{Data acquisition campaigns}
\label{subsec:data_acquisition_campaign}
\begin{figure}[htb]
    \centering
    \includegraphics[width=1.0\textwidth]{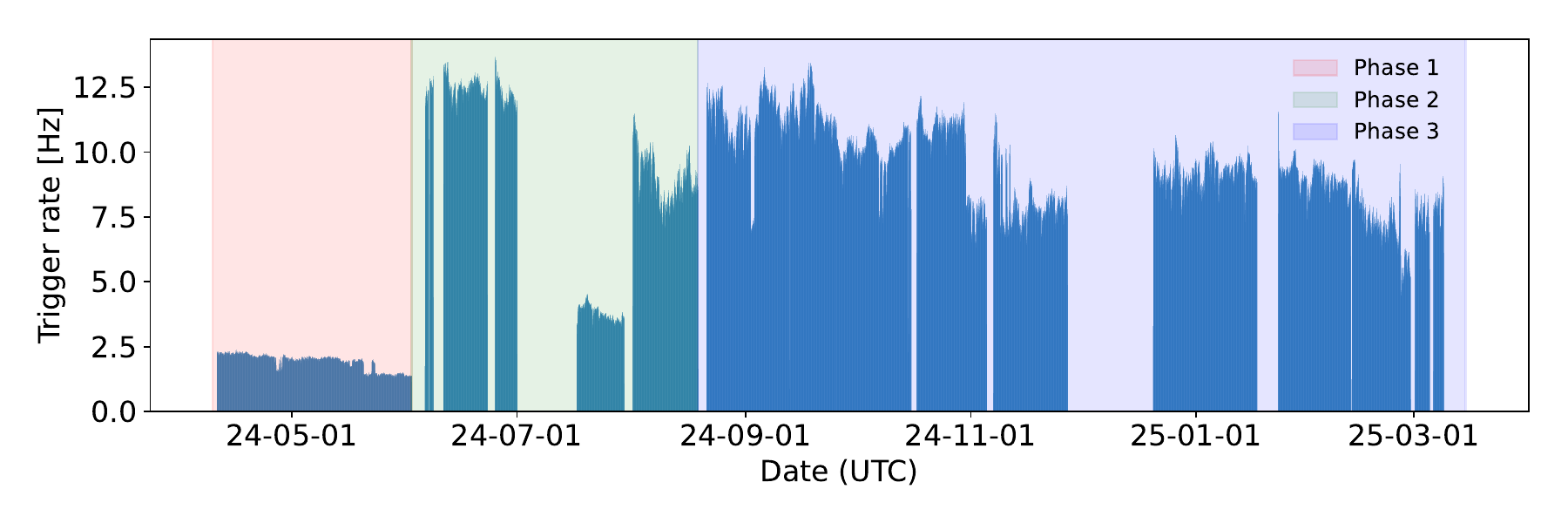}
    \caption{Trigger rate as a function of time. The three data acquisition phases, corresponding to different configurations of the PMT voltage settings, are represented by the colored intervals. PMT current trips, where the overcurrent threshold of the HV board is surpassed, result in sudden reductions in the trigger rate. 
    \label{fig:trigger_rate}}
\end{figure}

The muon data acquisition used for this study is divided into three phases, with the first phase commencing in April 2024. In this phase, the HV of all PMTs was set to 1500\,V, and the trigger rate at which the digitiser records a multi-panel event was approximately 2\,Hz. In the second phase, to improve the separation of muon events from noise, the absolute value of the voltage of all channels is increased to 1600\,V as of the 5th of June, 2024. In this second phase, the trigger rate is approximately 12\,Hz. However, due to voltage issues in some detector panels, the trigger rate dropped to approximately 4\,Hz. In the third phase, from the 21st of August 2024, the gains of different PMTs are matched; PMT gains are found to vary around 20\%. The trigger rate is approximately 10\,Hz. The trigger rate as a function of time is shown in Fig.~\ref{fig:trigger_rate}. All acquisition periods are used in the measurement as the muon rate is found to be independent of the HV setting, as shown in Table~\ref{tab:muon_flux_results} and Fig.~\ref{fig:Weekly_flux}. 
The operating voltages and currents of the PMTs are collected and stored in the Conditions Database, which can be connected to and visualised using the open source software tool Grafana~\cite{grafana}.

\subsection{Event selection}
\label{subsec:event_selection}
Muons traversing the 50\,mm thick EJ200 plastic scintillator are expected to lose a minimum energy of 10\,MeV based on the stopping power of a minimum ionising particle in polyvinyl toluene of 1.956 $\textrm{MeV/(g/cm}^2$). The energy loss follows a Landau distribution~\cite{Kolbig:1983uf}, parameterised as Landau($x$,\:$x_0$,\:$\xi$), where $x_0$ and $\xi$ are the location and scale parameters, respectively. For any muon hit, the light collected by each PMT depends on the distance between the hit and the PMT, following an exponentially decaying profile due to the light attenuation experienced by the signal in the scintillator. To cancel this position dependence, the so-called combined charge variable is used, defined as the geometric mean of the charges from the two PMTs,
\begin{equation}
Q_\mathrm{com} = \sqrt{Q_A Q_B}\,\propto\, \sqrt{\frac{I_0}{2} e^{-(\frac{L}{2}+x)/\lambda} \cdot \frac{I_0}{2} e^{-(\frac{L}{2}-x)/\lambda}} = \frac{I_0}{2}e^{-\frac{L}{2\lambda}}.
\label{eq:Qcom}
\end{equation}
where $Q_A$ and $Q_B$ represent the integrated charges from two PMTs coupled to the same panel, $L$ is the length of the panel, $x$ is the muon hit position along the panel, $\lambda$ is the light attenuation length, and $I_0$ is the original light signal intensity. 

\begin{figure}[htb]
    \centering
    \includegraphics[trim={0cm 0cm 0cm 1.2cm}, clip, width=1.0\textwidth]{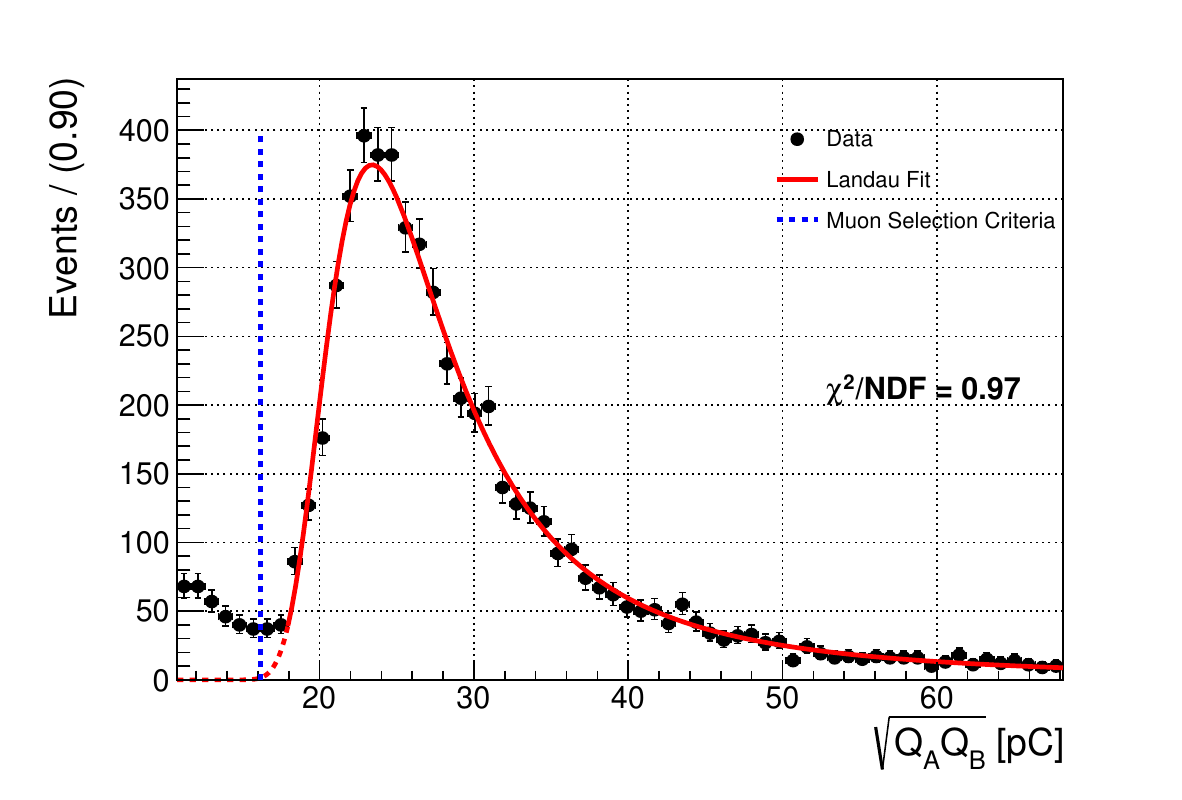}
    \caption{Landau fit to the combined charge spectrum. The blue line represents the lower bound measured using the Landau fit parameters.
    \label{fig:landau_fit}}
\end{figure}

Backgrounds in muon detection arise from gamma rays emitted by $^{40}$K and from the natural decay chains of $^{238}$U and $^{232}$Th present in the rock and construction materials~\cite{Keblbeck:2024mtb, Haffke:2011fp,Fearon:2024wkw}. Background energy deposits are typically below 3\,MeV, well below the 10 MeV deposit expected from the muon signal. A lower threshold on the charge is applied to separate signals from the background. A Landau fit is performed on a combined charge interval starting from a local minimum of the distribution. The selection criterion is defined using the Landau fit parameters, specifically the lower threshold of $x_0 - 3\xi$. Figure~\ref{fig:landau_fit} illustrates an example of this selection criterion where the background contributes below the 16.14\,pC threshold.

\begin{figure}[htb]
    \centering
    \includegraphics[width=0.49\textwidth]{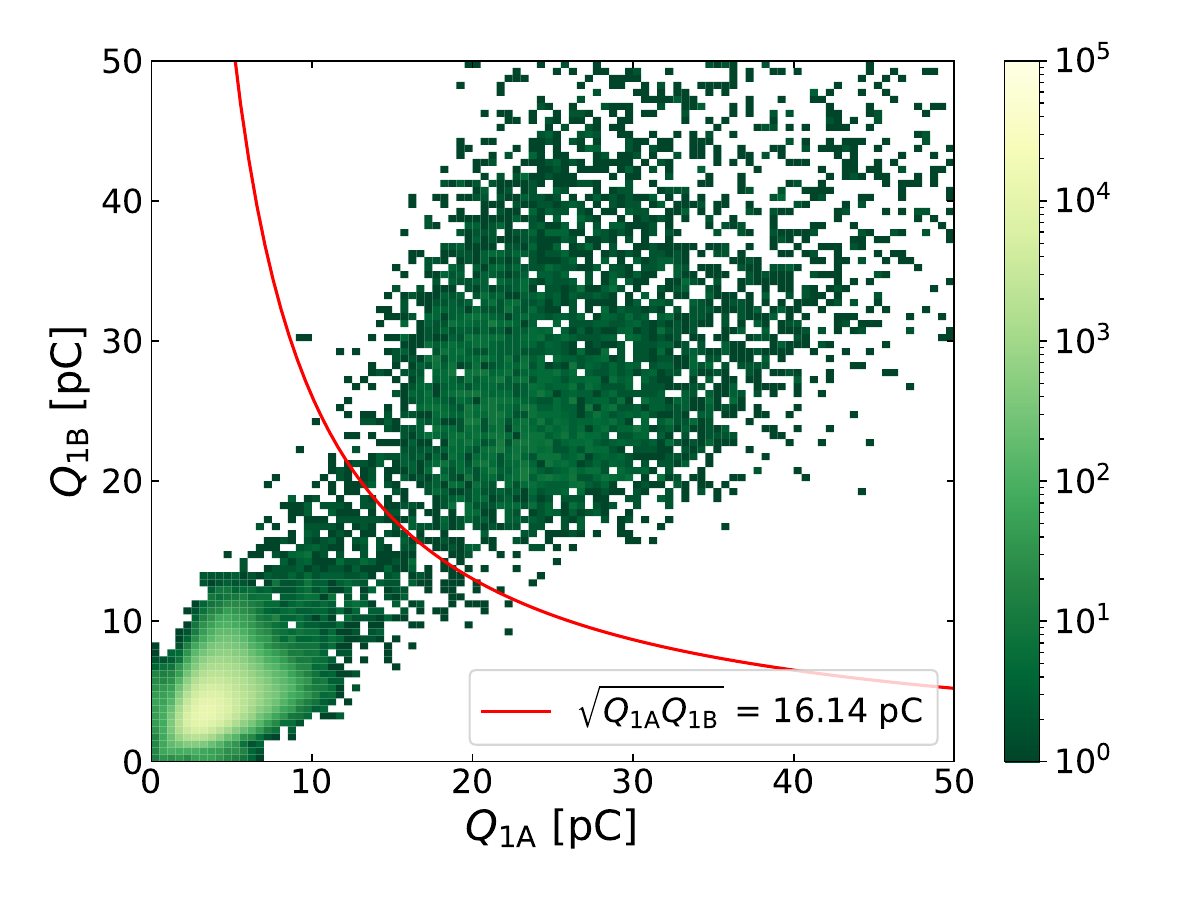}
    \includegraphics[width=0.49\textwidth]{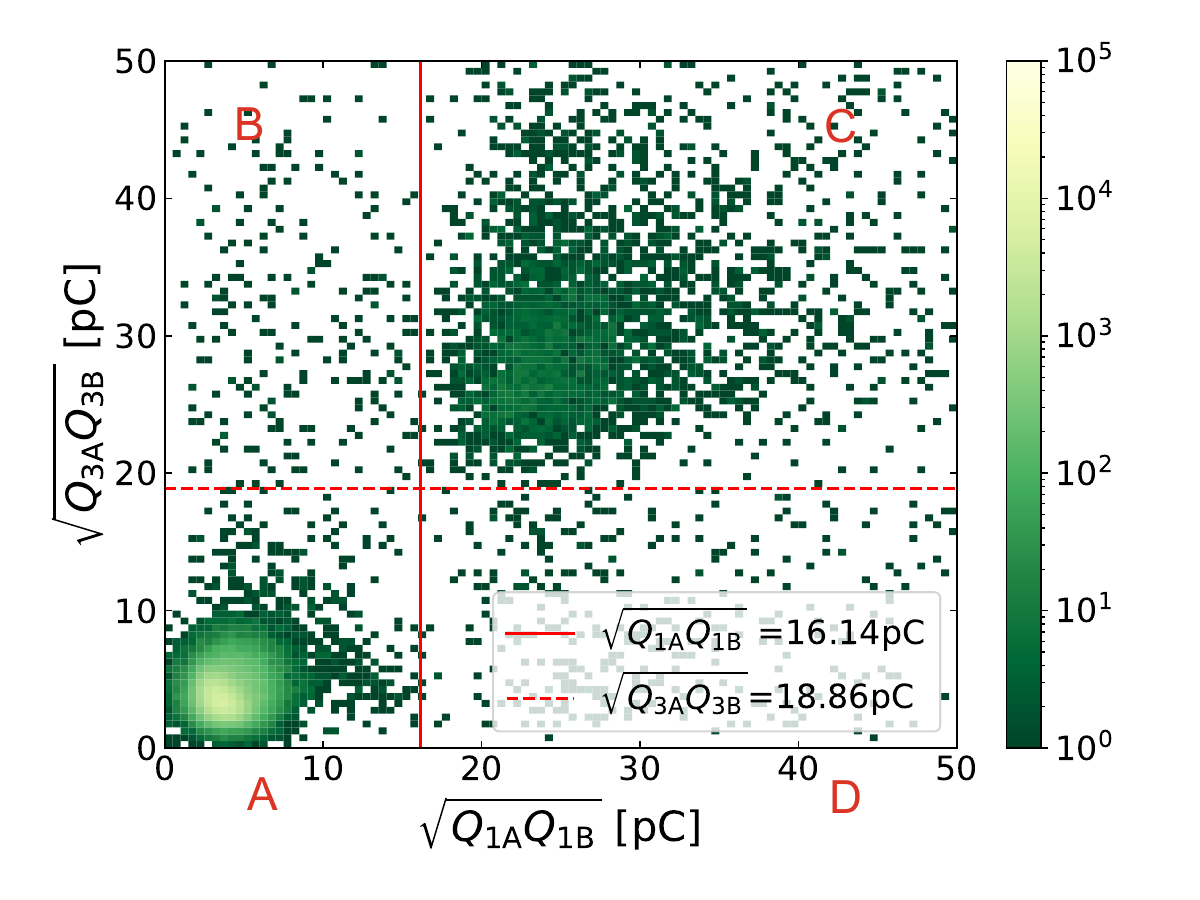}
    \caption{Left: Two-dimensional histogram showing the charge distribution between the two PMT channels from the same muon detector panel, and the selection criteria. The red curve represents the combined charge selection criterion illustrated in Fig.~\ref{fig:landau_fit}. Events below this line are dominated by the ambient $\gamma$ background. Right: Two-dimensional histogram of the combined charge distribution from a top and bottom muon panel, and the muon selection criteria from individual panels. The bin widths are 0.5\,pC, and the bin content, representing the number of events in each bin, is shown in a logarithmic scale from 1 to $10^5$. The two red lines indicate the threshold values used to discriminate against the background. This is located in the bottom-left region and includes $\gamma$ radiation. Those events in the top-right correspond to muons, while events in the top-left and bottom-right areas are from muons not traversing the entire 5 cm thickness of the scintillator. Geometrically speaking, these events are configured as types D and B in Fig.~\ref{fig:event_vis}, while events of type A are found in the bottom-left area together with the non-muon background.
    \label{fig:Panel_ChargeSpec}}
\end{figure}

A two-dimensional charge distribution of the two channels from the same panel is shown in Fig.~\ref{fig:Panel_ChargeSpec} (left). First, muon hits are identified for each panel by combining signals from its two channels and requiring that the combined charge exceeds a selection threshold. Then an event is categorised as a muon signal when at least one top and one bottom panel from the same telescope, each passing the panel-level selection, are triggered. Figure~\ref{fig:Panel_ChargeSpec} (right) shows the combined charge distribution from a vertically aligned pair of panels. The Landau boundary value chosen to define signal events introduces an angular dependence on the acceptance of muons. This is henceforth referred to as an {\it edge effect} and will be investigated in Sec.~\ref{subsec:geometric_acceptance_sys}.

\begin{figure}[htb]
    \centering
    \includegraphics[width=0.55\textwidth]{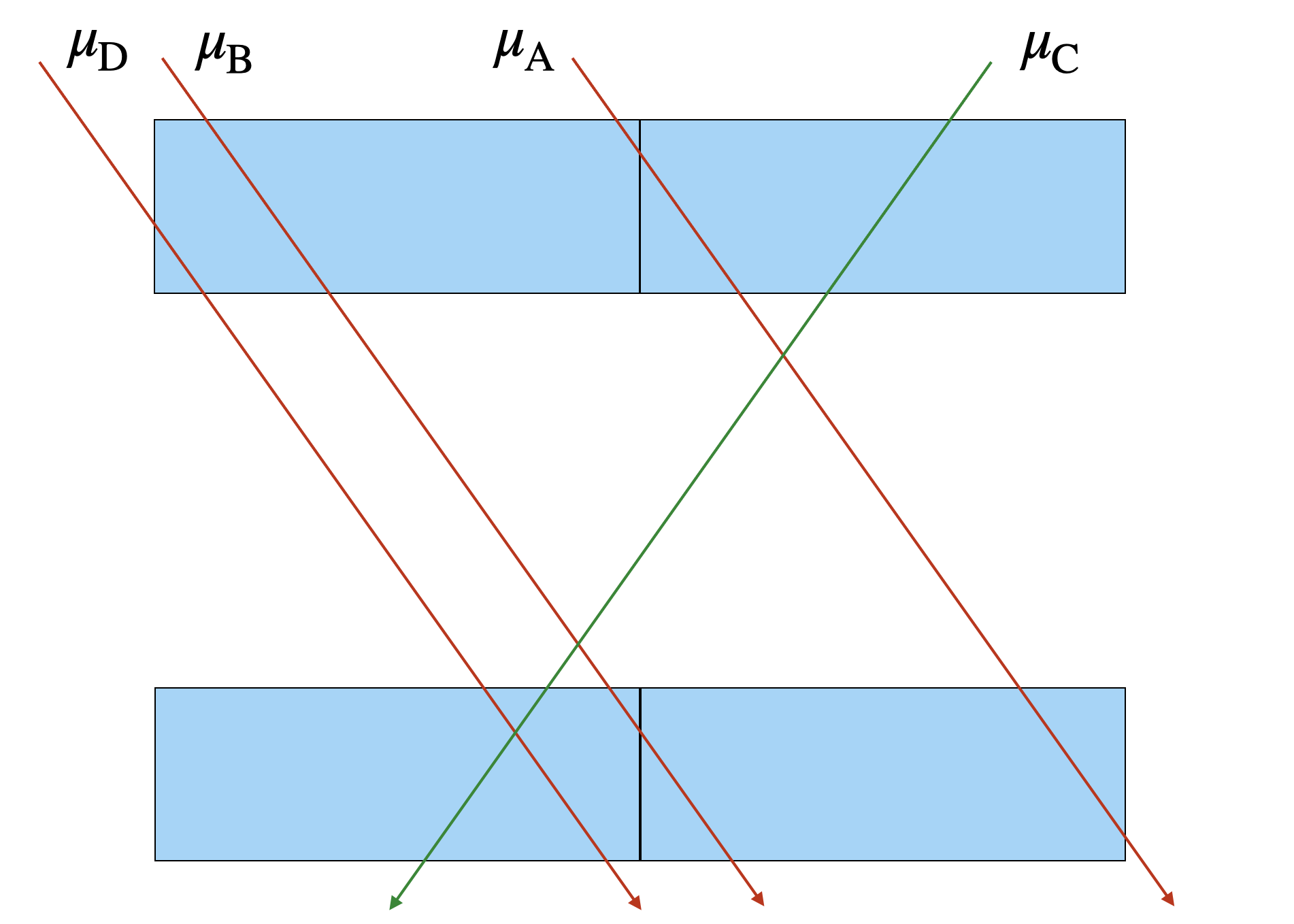}
    \caption{Qualitative representation of true muon event geometries detected by the muon telescope. Events are labelled according to the selection regions in Fig.~\ref{fig:Panel_ChargeSpec} (right). Events in region C, traversing the full thickness of the scintillator, are considered signal.}
    \label{fig:event_vis}
\end{figure}

\begin{figure}[htb]
    \centering
    \includegraphics[width=1.0\textwidth]{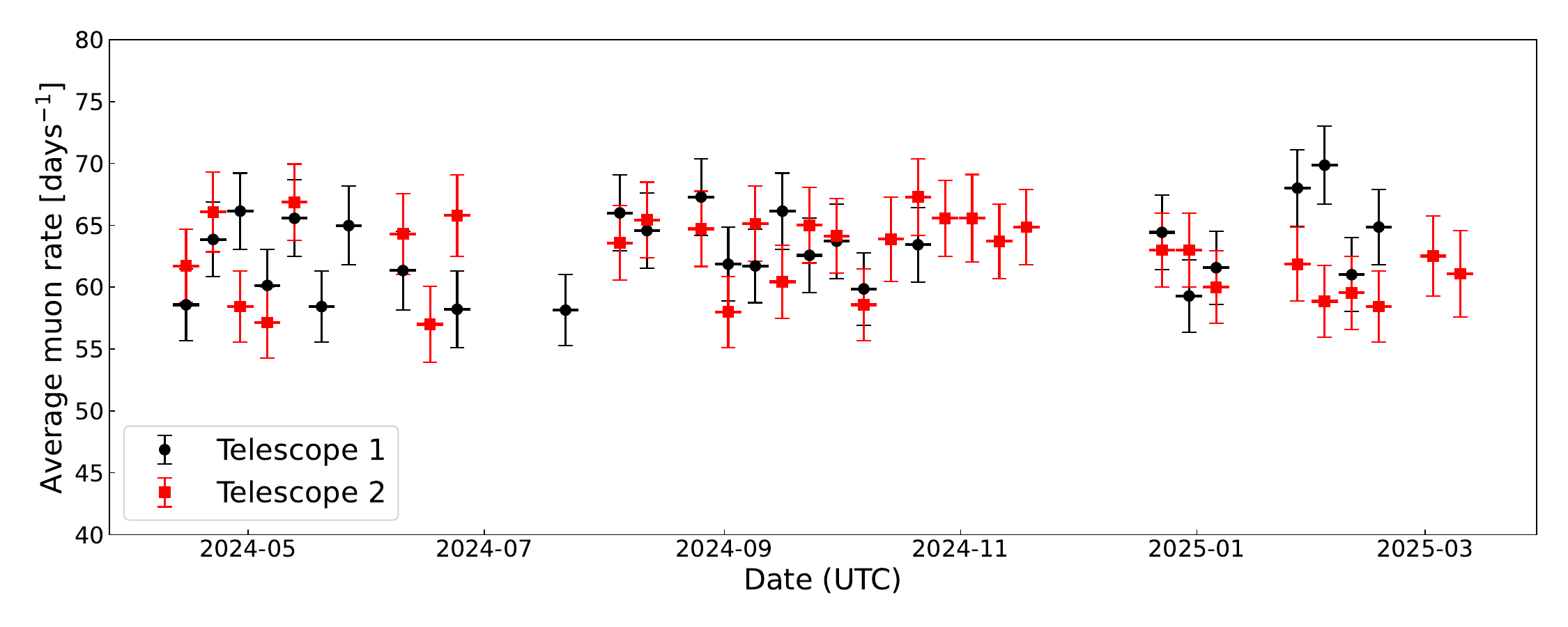}
    \caption{Muon rate measured at SUPL as a function of time. The daily rate is averaged over a week.
    \label{fig:Weekly_flux}}
\end{figure}

\begin{figure}[htb]
    \centering
    \includegraphics[width=1.0\textwidth]{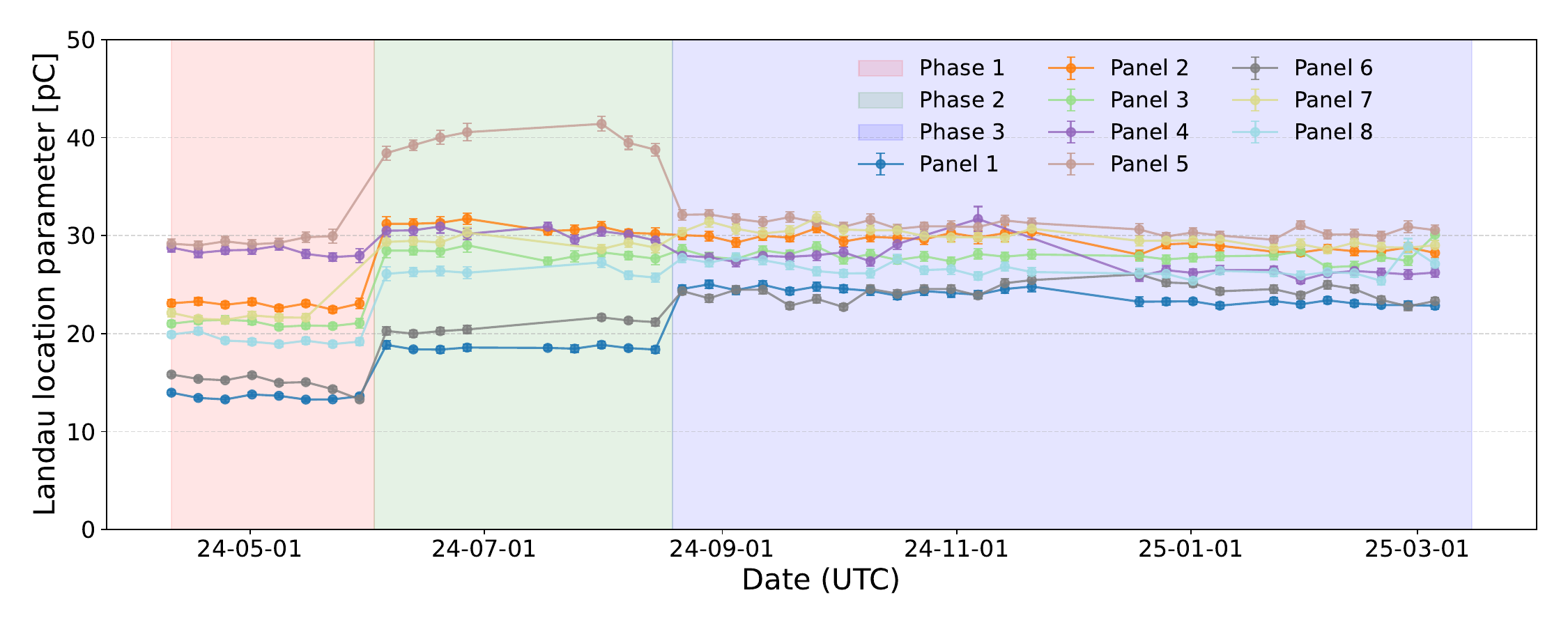}
    \caption{Evolution of the gain by monitoring the location parameter of the Landau fit to the combined charge of each panel.
    \label{fig:Landau_MPV_Evolution}}
\end{figure}

By counting the muons detected by the two telescopes, we can get the average muon rate as a function of time, as shown in Fig.~\ref{fig:Weekly_flux}. The data affected by PMT voltage faults is removed from the analysis of the corresponding telescope. The muon rate is stable throughout the data-taking period, although the gain of the muon panels changes as monitored by the location parameter and illustrated in Fig.~\ref{fig:Landau_MPV_Evolution}. Muon counts, uptime, and daily rates from the three data-taking periods are summarised in Table~\ref{tab:muon_flux_results}.

\begin{table}[htbp]
    \centering
    \caption{Muon counts, uptime, and daily rates measured under different voltages by the two muon telescopes.}
    \label{tab:muon_flux_results}
    \resizebox{\textwidth}{!}{
       \begin{tabular}{lcccccc}
        \toprule
        \multirow{2}{*}{Voltage} & \multicolumn{3}{c}{Telescope 1} & \multicolumn{3}{c}{Telescope 2} \\
        & Counts & Uptime [d] & Rate [d$^{-1}$] & Counts & Uptime [d] & Rate [d$^{-1}$] \\
        \midrule
        1500 V       & 3189  & 51  & $62.5 \pm 1.1$ & 2311  & 37  & $62.5 \pm 1.3$ \\
        1600 V       & 3000  & 49  & $61.2 \pm 1.1$ & 2291  & 36  & $63.6 \pm 1.3$ \\
        Gain matched & 8812  & 139 & $63.4 \pm 0.7$ & 10079 & 161 & $62.5 \pm 0.6$ \\ \hline
        Total & 15001  & 239 & $62.8 \pm 0.5$ & 14681 & 234 & $62.7 \pm 0.5$ \\
        \bottomrule
        \end{tabular}
        }
\end{table}

\section{Detector Performance}
\label{sec:detector_performance}
To convert the measured muon rate to a flux measurement, we must account for detector and acceptance effects. 
Ultimately, the final muon flux measurement is determined as
\begin{equation}
\label{eq:muon_flux_final}
    f = \frac{f_{\rm raw}}{\varepsilon\;\alpha}
\end{equation}
where $f_{\rm raw}$ is the raw muon flux measured by the detector as the muon rate divided by the detector surface, $\varepsilon$ is the muon detection efficiency of the telescope and $\alpha$ its geometrical acceptance. Data collected above ground and simulations are used to quantify $\varepsilon$ and $\alpha$ as described in the following two sections.

\subsection{Muon detection efficiency}
\label{subsec:detection_efficiency}

\begin{figure}[htb]
    \centering
    \includegraphics[width=0.55\textwidth]{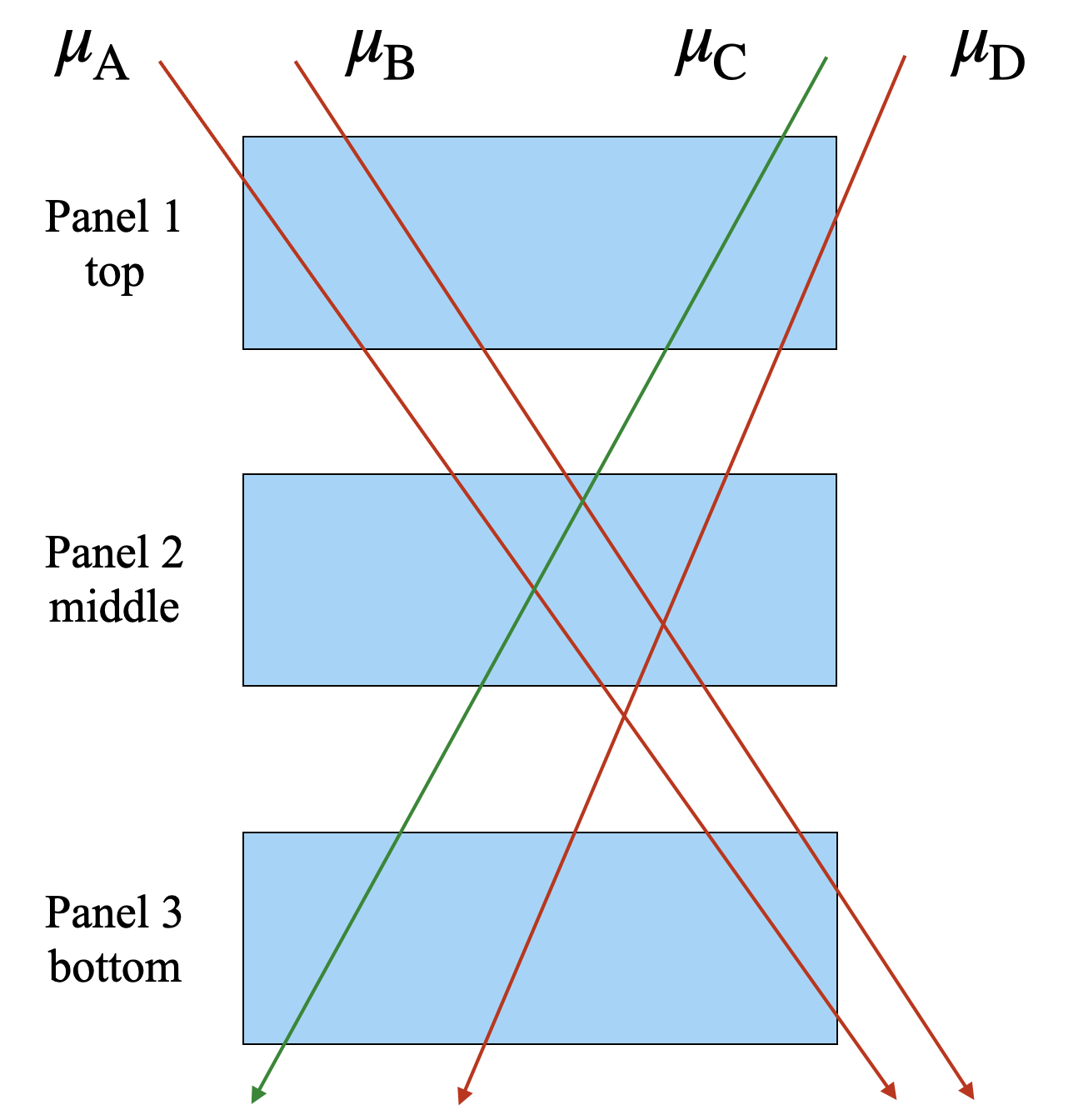}
    \caption{Qualitative representation of true muon event geometries relevant for the measurement of the single-panel muon detection efficiency. Events are labelled according to the selection regions in Fig.~\ref{fig:Edge_Effect}. Events in region C, traversing the full thickness of the scintillator, are considered signal.}
    \label{fig:event_vis_trigger_efficiency}
\end{figure}

To measure the detection efficiency of the muon telescope, the single-panel detection efficiency is investigated for muons crossing the entire 5\,cm thickness of the scintillator. This is performed using data collected above ground, where the muon flux is significantly higher. Three panels are arranged vertically, as illustrated in Fig.~\ref{fig:event_vis_trigger_efficiency}, with a middle layer, named panel 2, positioned between the top and bottom panels, 1 and 3, respectively. The vertical separation between the top and middle panels is 6.9\,cm, while the middle and bottom panels are separated by 8.5\,cm. The time difference between signals recorded by the two PMTs of the same panel is expected to be less than approximately 20 ns. A pair-level trigger is formed by applying a logical AND to the two signals, aiming at filtering out low-energy noise events, and using a channel trigger gate width of 100 ns. The board-level trigger is then generated with a logical OR of the three pair triggers.

\begin{figure}[htb]
    \centering
    \includegraphics[width=1.0\linewidth]{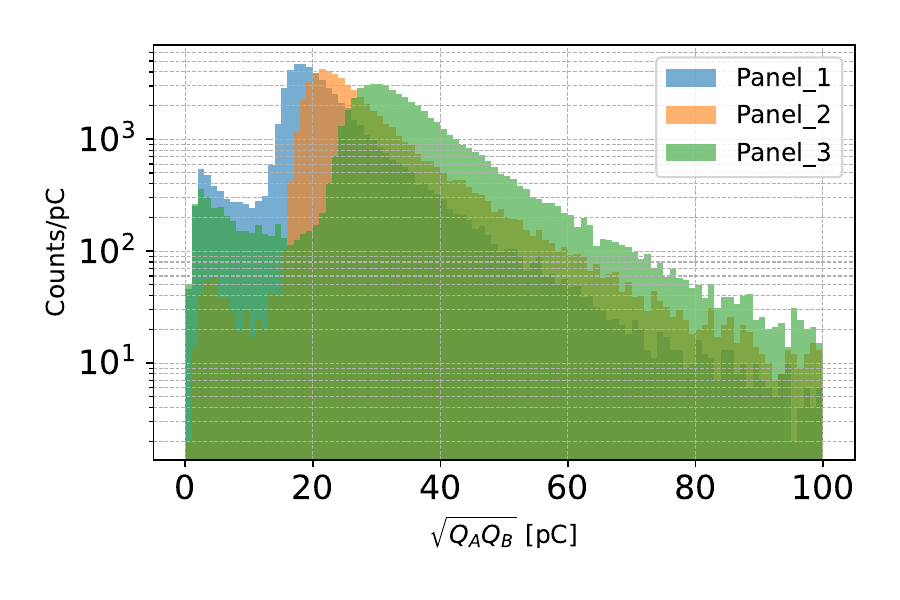}
    \caption{The combined charge distributions of the three panels for events passing the software top--bottom coincidence selection.}
    \label{fig:Tripple_ChargeSpec_TB}
\end{figure}

\begin{figure}[!htbp]
    \centering
    \includegraphics[width=0.65\linewidth]{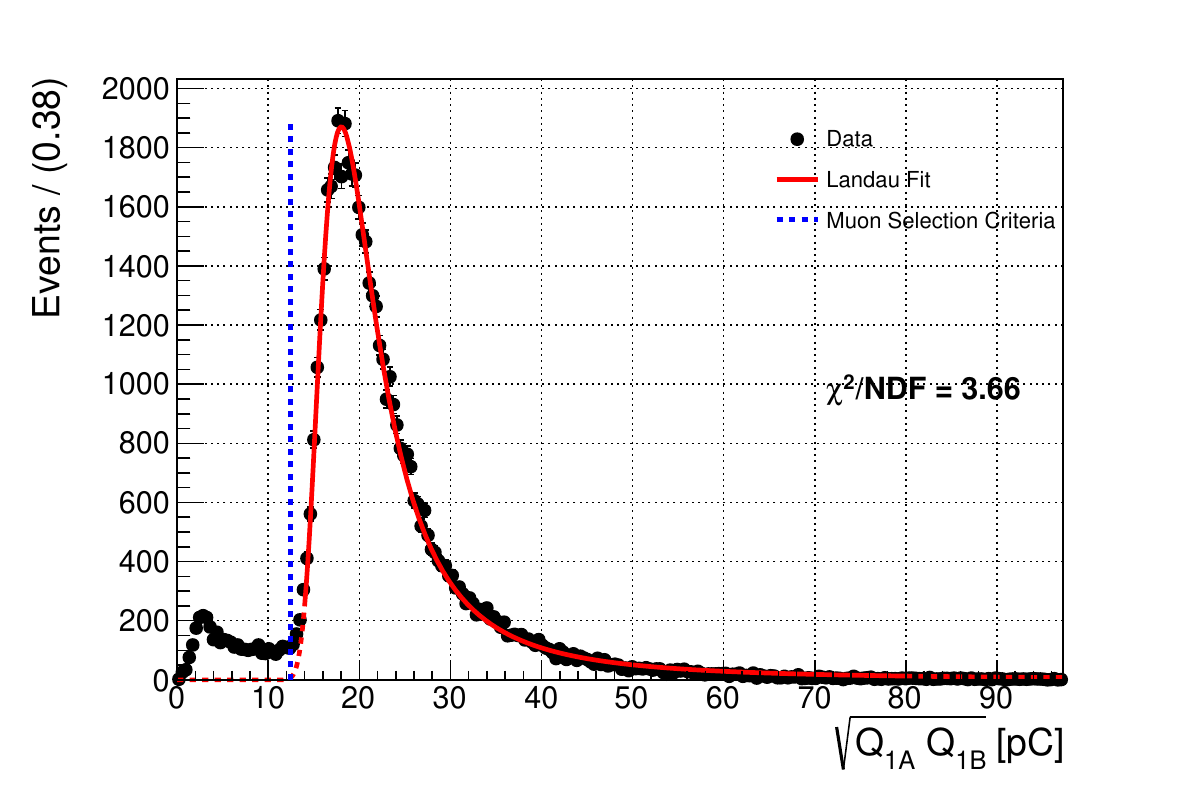}
    \includegraphics[width=0.65\linewidth]{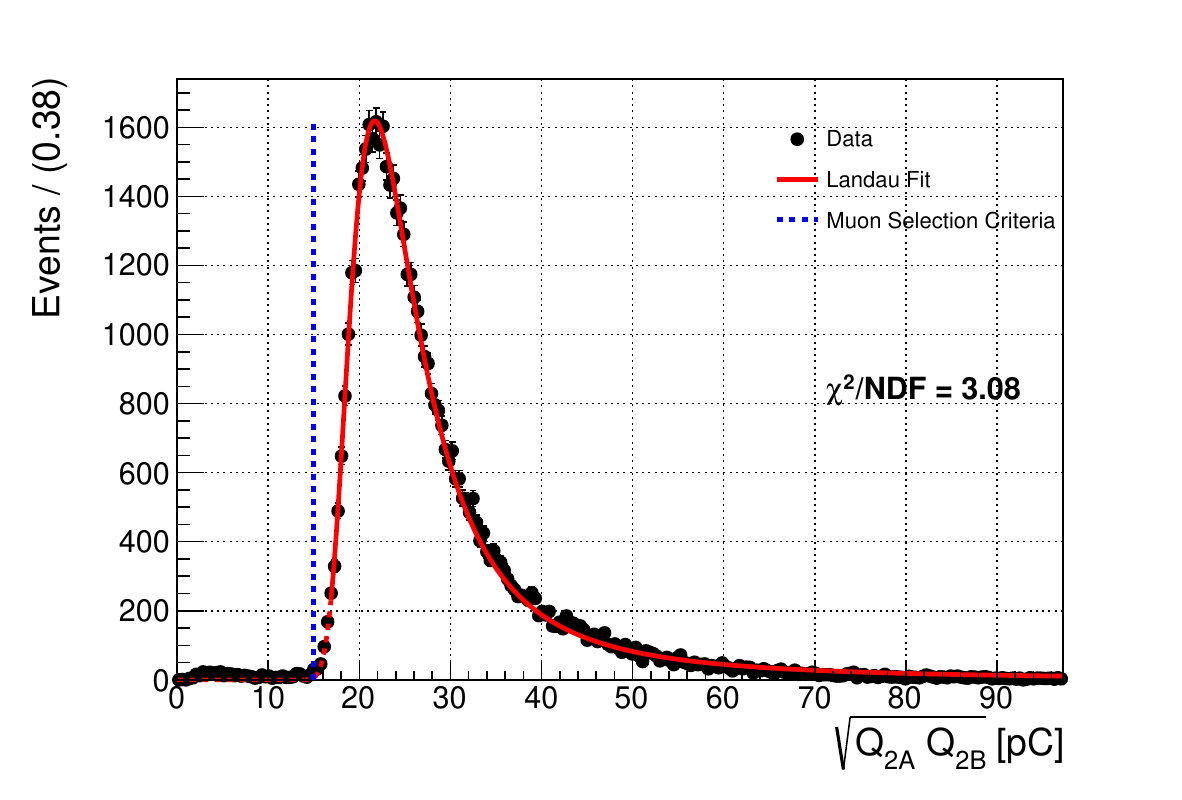}
    \includegraphics[width=0.65\linewidth]{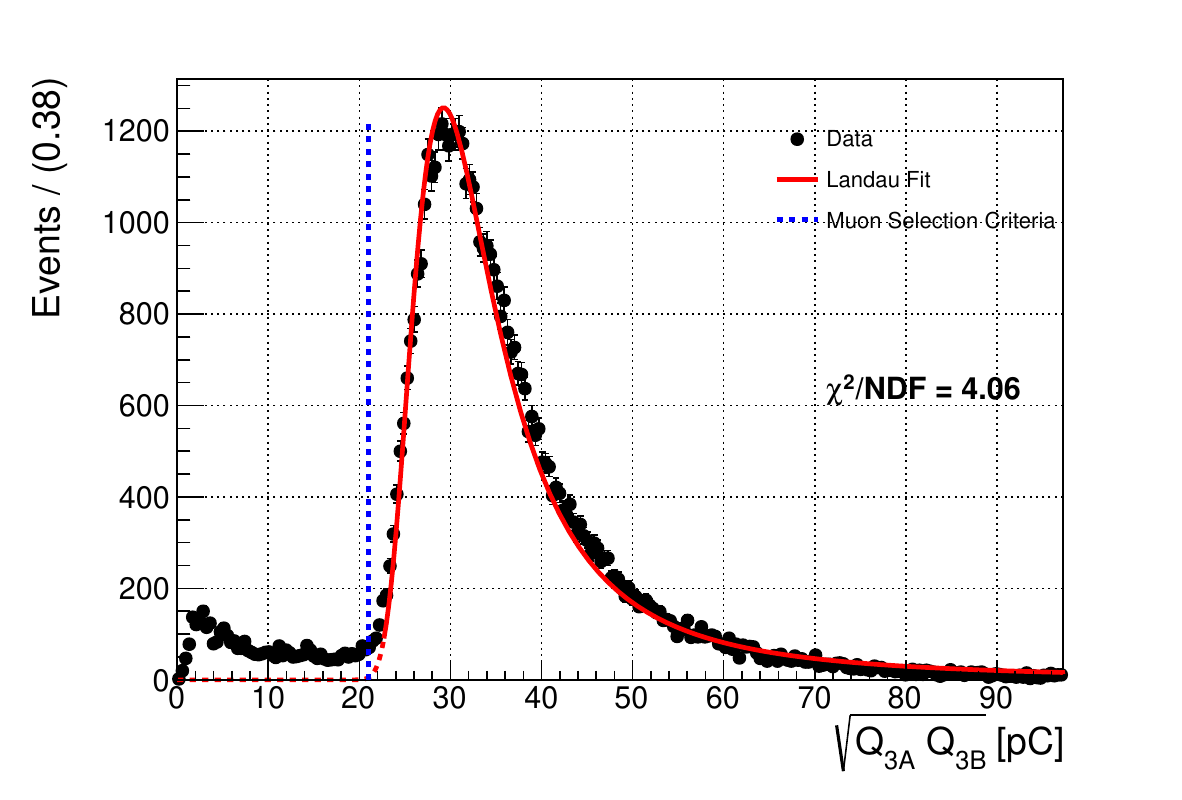}
    \caption{Landau fit to the combined charge spectra of the three panels used to measure the muon detection efficiency. Events are selected offline by requiring coincidences between the top and bottom panels only.}
    \label{fig:Tripple_ChargeSpec_landau_fit}
\end{figure}

Choosing from the 10-minute run, two types of events are counted offline: those requiring an offline coincidence between the top and bottom panels only, and those with a middle layer coincidence too. The combined charge distribution of the three panels, for events passing the offline top--bottom coincidence, is shown in Fig.~\ref{fig:Tripple_ChargeSpec_TB} and in Fig.~\ref{fig:Tripple_ChargeSpec_landau_fit}. Two-dimensional distributions are shown in Fig.~\ref{fig:Edge_Effect} for events triggered only by the top and bottom panels (left) and by all three panels (right).

An offline selection criterion based on a combined charge threshold for muon candidates is applied to further remove the ambient gamma background. As in the analysis described in Sec.~\ref {subsec:event_selection}, we fit a Landau distribution to the combined charge spectra of the three panels as shown in Fig.~\ref{fig:Tripple_ChargeSpec_landau_fit}, to guarantee a background-free muon sample. The blue dashed line indicates the muon selection criteria for each panel. The Landau fit is performed on all three distributions only for events triggered by the top and bottom panels, as signal events above threshold would not considerably change for triple coincidences, as evident from the two-dimensional distributions in Fig.~\ref{fig:Edge_Effect}. After the fit is performed and events below threshold are rejected, the ratio of triple coincidences over double ones is measured to determine the single-panel muon detection efficiency as

\begin{equation}
\label{eq:muon_efficiency}
    \varepsilon_\text{2} = \frac{N_\text{1,2,3}}{N_\text{1,3}},
\end{equation}
where $N_\text{1,2,3}$ is the number of events selected offline by requiring coincidence between all panels, with the charge in each panel exceeding the threshold, while $N_\text{1,3}$ is the number of events selected using only the top--bottom coincidence and the same charge selection criterion.

\subsection{Detection efficiency systematic uncertainty}
\label{subsec:detection_efficiency_sys}

\begin{figure}[!htbp]
    \centering
    \includegraphics[width=1.0\linewidth]{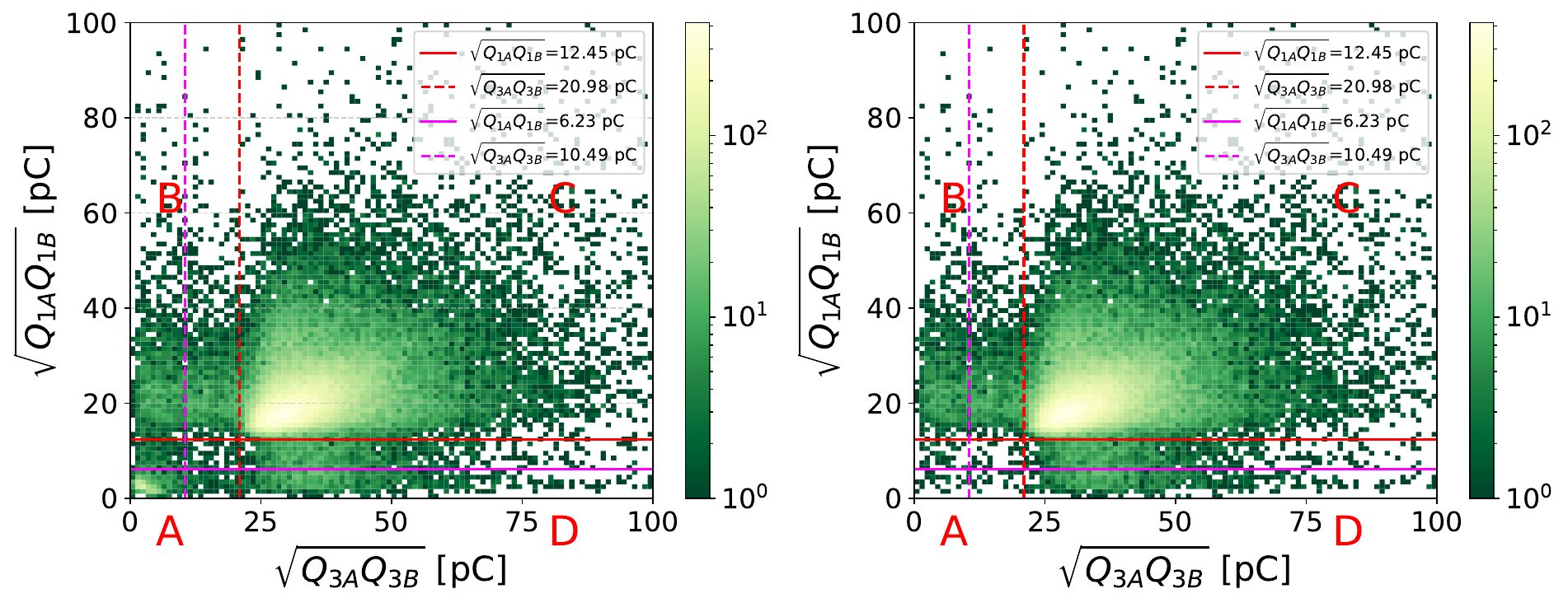}
    \caption{Two-dimensional distribution of the combined charge for the bottom and top panels. The red and magenta lines are the nominal and modified event selection criteria, respectively. Left: Events are selected offline by requiring coincidences between the top and bottom panels only. Right: Events are selected offline by requiring coincidences between all three panels.}
    \label{fig:Edge_Effect}
\end{figure}

The angle of incidence of incoming muons can affect the reconstruction efficiency, as light reflection patterns inside the central panel vary. As shown in Fig.~\ref{fig:event_vis_trigger_efficiency}, however, the angle of incidence can be controlled using the energy deposition in the top and bottom panels, where increasingly larger angles of incidence are correlated with smaller energy deposits. Energy deposits in the top and bottom panels are used as a handle to estimate this effect via the Landau selection criterion which is relaxed to half its nominal value, before re-evaluating the efficiency. The criteria used, nominal and modified, respectively corresponding to $x_0 - 3\xi$ and $(x_0 - 3\xi)/2$, are illustrated in Fig.~\ref{fig:Edge_Effect}. It is important to notice that relaxing the Landau criterion is expected to also increase non-muon background. Nonetheless, the uncertainty is conservatively taken as a symmetric effect on the efficiency, assuming this to be maximal for vertical muons which cannot be exclusively selected altering the Landau criterion.

The total number of events in each region is reported in Table~\ref{tab:Edge_effect_contribution}, where muons traversing the full scintillator thickness correspond to region C. After comparing event counts using the modified criteria to those from the nominal measurement, the final result on the single-panel muon detection efficiency is

\begin{equation}
\label{eq:muon_detection_efficiency_meas}
    \varepsilon_{\rm 2} = 99.42\pm0.03_\mathrm{stat}\pm0.23_\mathrm{sys}\,[\%].
\end{equation}

\begin{table}[htb]
    \centering
    \caption{Events triggered by all three panels ($1,2,3$) and top and bottom panels only ($1,3$) in different regions with nominal and modified event selection criteria.}
    \label{tab:Edge_effect_contribution}
    \resizebox{\textwidth}{!}{
    \begin{tabular}{ccccccc}
        \toprule
         Software coincidence &Event selection& A & B & C & D &Total\\
        \hline
         \multirow{2}{*}{Panels 1,2,3}
        & Nominal& 122&2308&50531&2374&55335\\
        &Modified& 39&1136&53279&1082&55536\\
        \hline
        \multirow{2}{*}{Panels 1,3}
        & Nominal&1128&2634&50827&2676&57265\\
        & Modified&790&1444&53713&1318&57265\\        
        \bottomrule
    \end{tabular}
    }
\end{table}

\subsection{Geometric acceptance}
\label{subsec:geometric_acceptance}
We determine the telescope geometric acceptance using simulated events. The geometric acceptance $\alpha$ appearing in Eq.~\ref{eq:muon_flux_final} can be written as:

\begin{equation}
\label{eq:acceptance}
    \alpha = \frac{\int_{\rm \Omega}\alpha_{\rm sim}(\eta, \phi)\:f_{\rm sim}(\eta, \phi)\:d{\rm \Omega}}{\int_{\rm \Omega} f_{\rm sim}(\eta, \phi)\:d{\rm \Omega}},
\end{equation}
where the integration is conducted over the full $2\pi$ hemisphere identified by the elevation and azimuth angles, respectively $\eta$ and $\phi$. In Eq.~\ref{eq:acceptance}, $f_{\rm sim}(\eta, \phi)$ represents the simulated muon flux at SUPL, while $\alpha_{\rm sim}(\eta, \phi)$ is the simulated acceptance of the muon telescope as a function of the angular coordinates. 

\begin{figure}[h]
    \centering
    \includegraphics[width=0.8\textwidth]{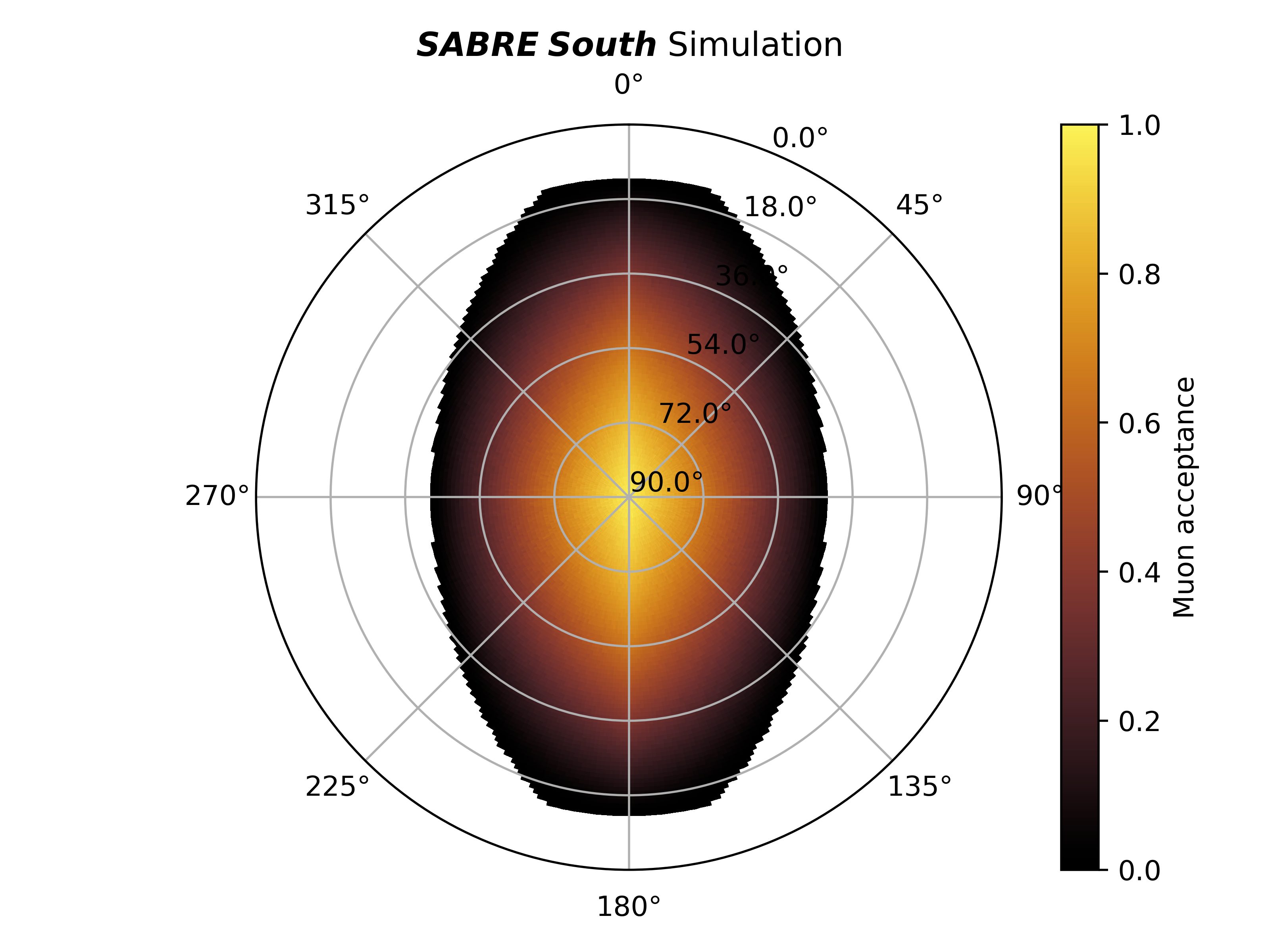}
    \caption{Geometrical acceptance of the muon detector arranged in telescope configuration as a function of the elevation and azimuth. Each angular bin, with a size of $\ang{1}\times\ang{1}$, contains the ratio between the accepted and generated muons in the angular range. The acceptance displayed here assumes an arbitrary orientation in which the detector length runs from North to South. The orientation of the detector in the laboratory does affect the integrated flux measurement and will be addressed as a systematic effect.}
   
    \label{fig:geo_acc}
\end{figure}

To measure $\alpha_{\rm sim}(\eta, \phi)$, a distribution of muon events is simulated using the \textsc{pyrate} framework~\cite{pyrate_2023} of the SABRE South experiment, together with the geometry package \textsc{pyvista}~\cite{pyvista} for the telescope geometry model. The volume of each top-layer panel is subdivided into a grid of $n_{\rm x} \times n_{\rm y} \times n_{\rm z} = 100 \times 100 \times 5$ points uniformly arranged across the top volume. For each point, 1000 muon events are simulated with a uniform angular density in the range $\eta\in[\ang{0},\ang{90})$ and $\phi\in[\ang{0},\ang{360})$. The ratio of events crossing the entire thickness of both top and bottom panels to all generated events is determined as a function of the angle in a fine angular grid and multiplied by an angular factor ${\sin(\eta)}$, which accounts for the effective area of the detector exposed to the flux. The geometric acceptance maps is shown in Fig.~\ref{fig:geo_acc}.

\begin{figure}[htb]
    \centering
    \includegraphics[width=0.49\textwidth]{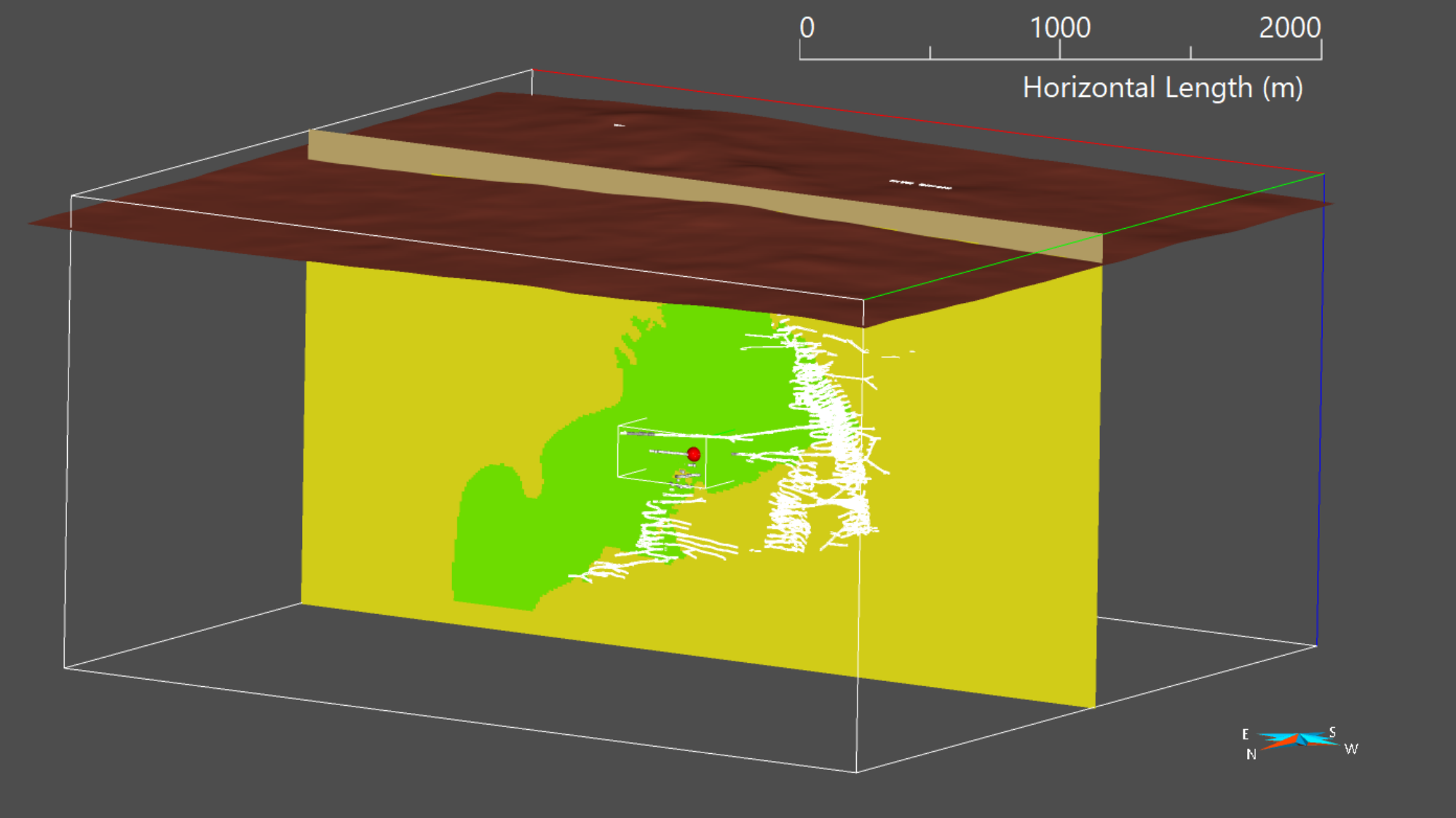}
    \includegraphics[width=0.501\textwidth]{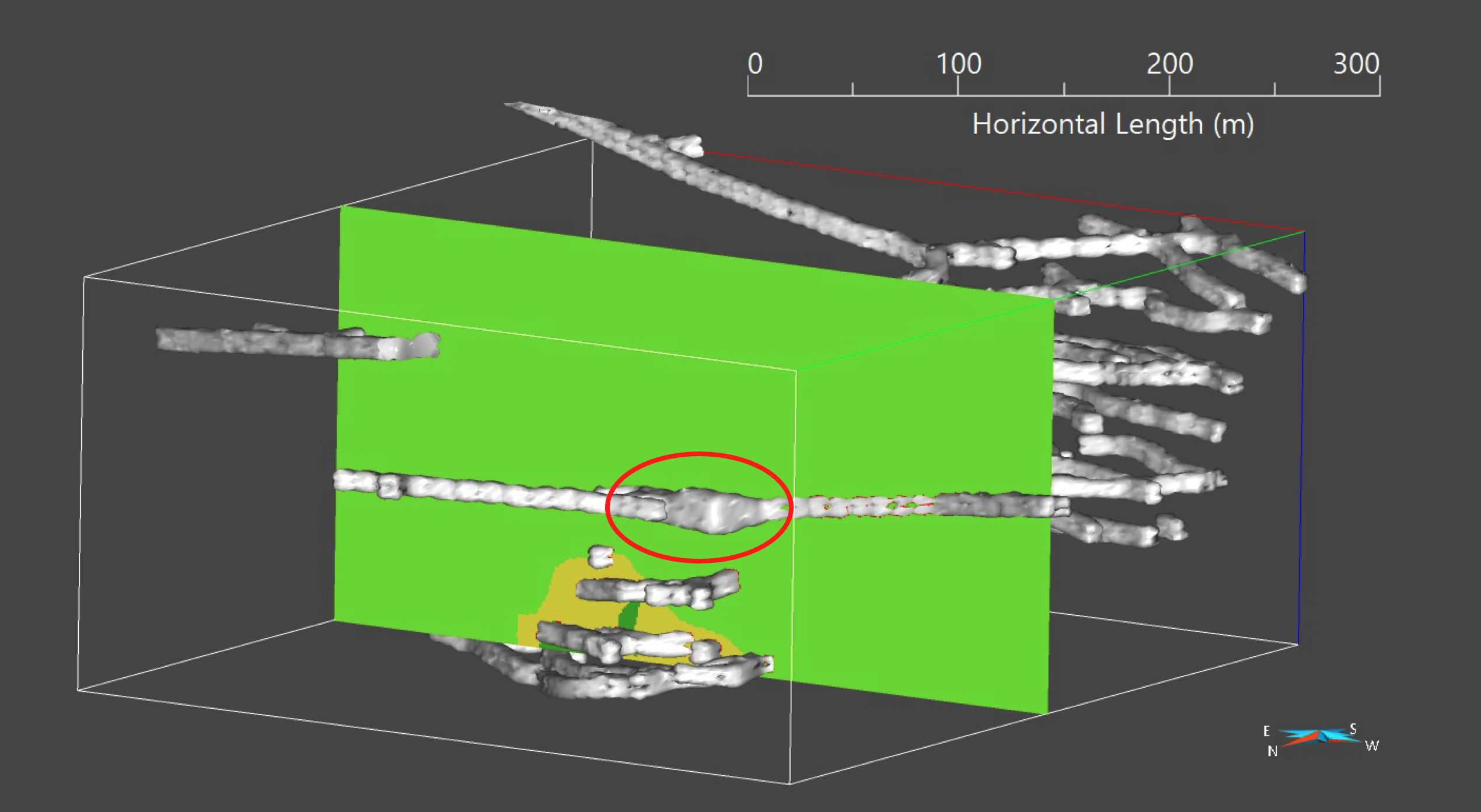}
    \caption{Left: Cross-sectional (East-to-West) view of the SUPL laboratory overburden. The basalt component is shown in green while the sandstone is in yellow. The visible portion of the tunnel decline structure is depicted in white, while the red dot indicates the location of the laboratory. Right: close-up detail of the cross-sectional overview on the left, where SUPL is shown at the centre of the picture and circled in red.}
    \label{fig:supl-cross-section}
\end{figure}

\begin{figure}[htb]
    \centering
    \includegraphics[width=0.49\textwidth]{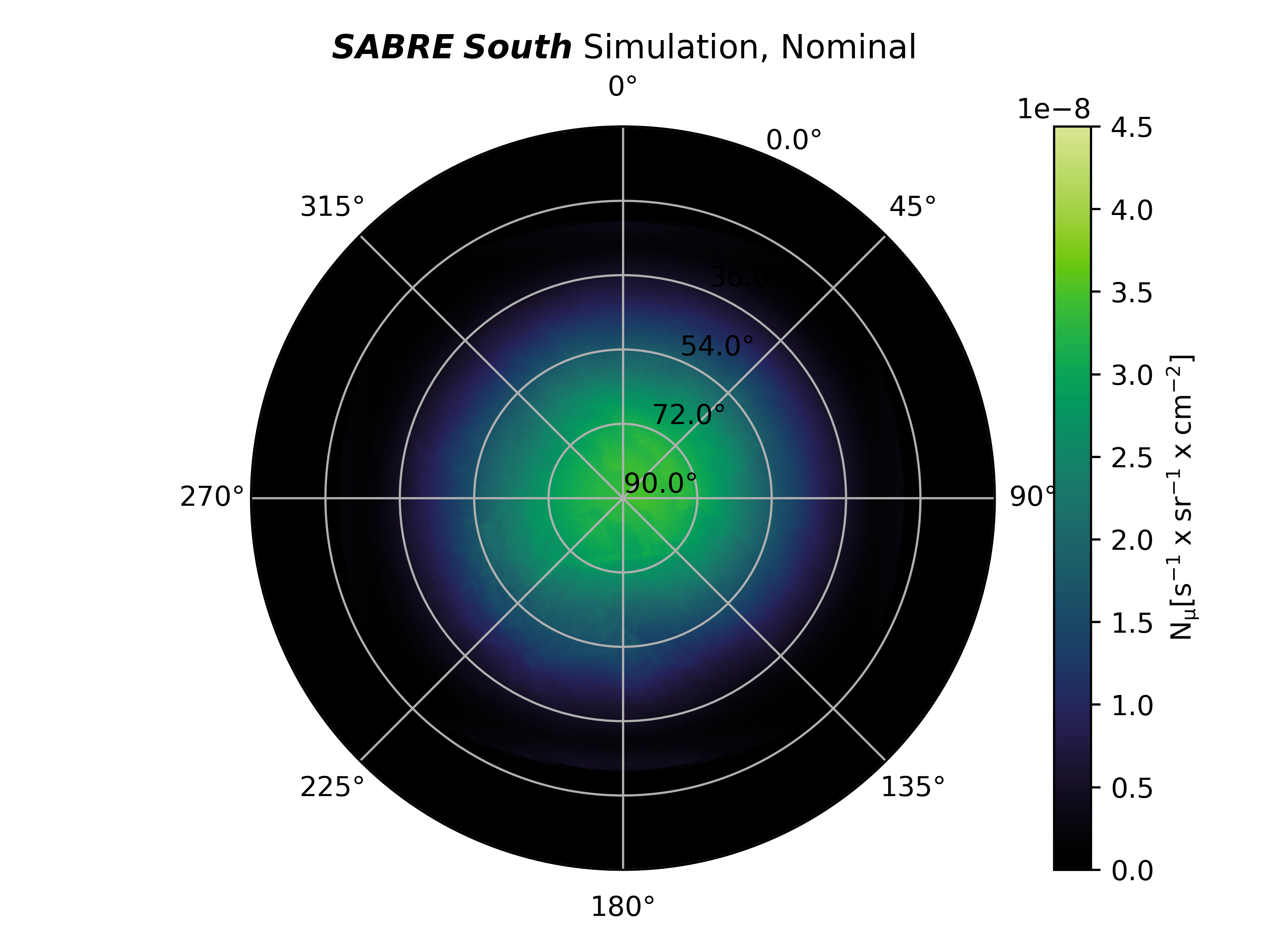}
    \includegraphics[width=0.49\textwidth]{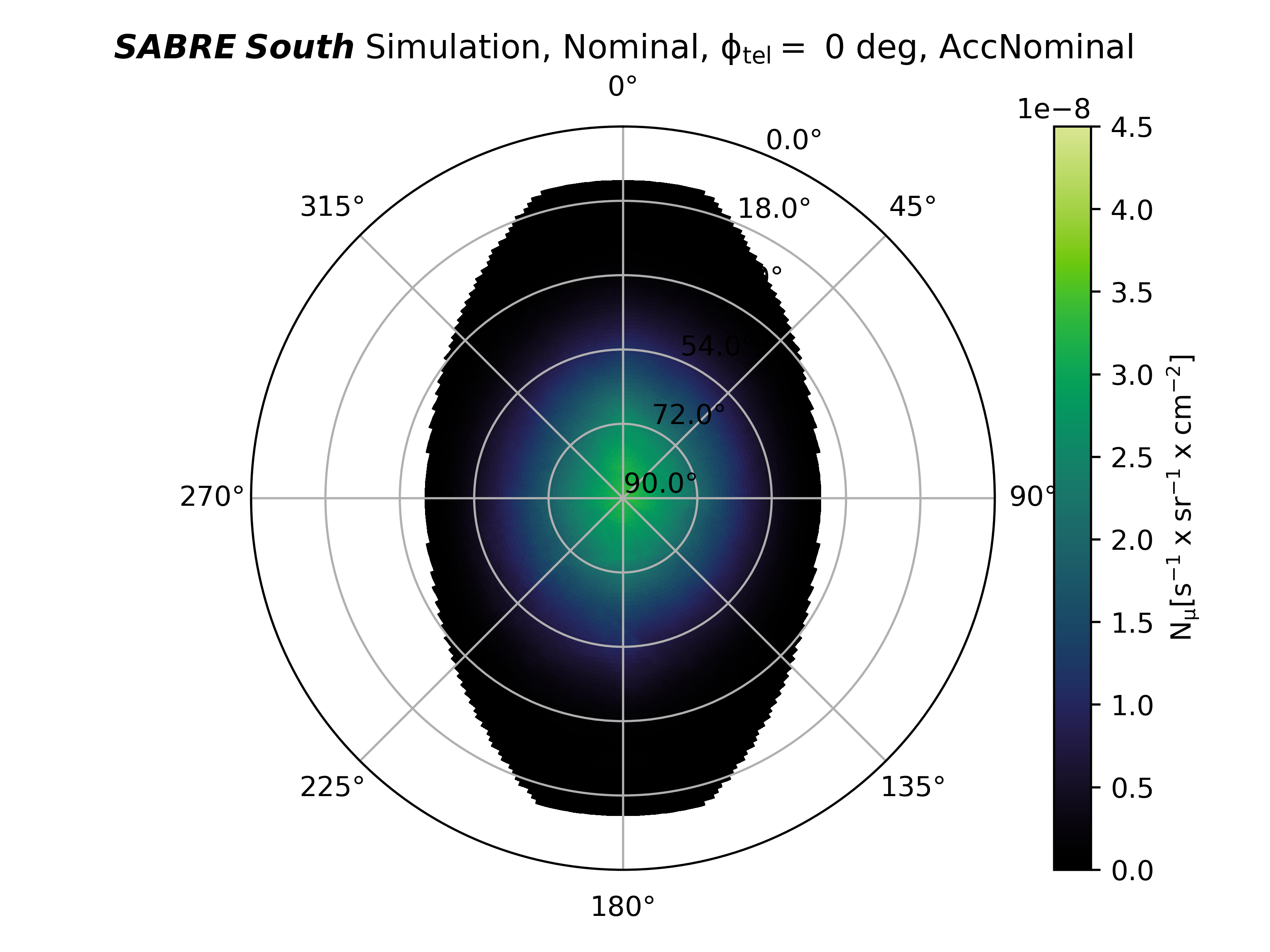}
    \caption{Simulated muon flux at SUPL as a function of the elevation and azimuth using \textsc{pumas}~\cite{pumas_1, pumas_2}. Left: simulated flux assuming a uniform angular acceptance of 100$\%$. Right: simulated flux where the detector acceptance has been taken into account, considering the nominal orientation of the detector with its length ranging from North to South.}
    \label{fig:pumas_flux}
\end{figure}

To obtain $f_{\rm sim}(\eta, \phi)$, we simulate the muon flux at SUPL using the \textsc{pumas} muon transport code~\cite{pumas_1} in backward mode. In this approach, muons are generated at the detector location and transported backwards through the overburden to the surface. This method dramatically improves computational efficiency compared to forward simulation, since all generated muons reach the detector by construction. The technique is analogous to importance sampling in Monte Carlo methods~\cite{pumas_2}, where events are sampled from a biased distribution to focus computational resources on relevant phase space regions. \textsc{pumas} computes the appropriate weight (Jacobian) to correct for this bias and recover the true muon flux distribution underground. This must be corrected by a factor dependent on the bias distribution and multiplied by the muon flux at the surface. The surface muon flux chosen for this computation is the updated Gaisser parameterisation from the model in Ref.~\cite{Guan:2015vja}. Muon transport is achieved through a series of steps, where the muon is transported through a discretised model of the mine. Voxels represent elements of constant density and material identity (lithology). These voxels have variable dimensions throughout the model, ranging from 10\,m at the edges to 1\,m in the vicinity of the SUPL laboratory and its immediate surroundings. The model is constructed using data from the Stawell Gold Mine, providing an accurate set of lithologies and material densities. The model includes, as dominant components of the overburden, basalt at $\rho_{\rm basalt} = 2.81\,\pm\,0.25\,{\rm g/cm^3}$ and sandstone belonging to the so-called Warrak formation of the St Arnaud group~\cite{warrak} at $\rho_{\rm Warrak} = 2.7\,\pm\,0.3\,{\rm g/cm^3}$. Uncertainties on these densities are provided by Stawell Gold Mine, and measured using extensive underground surveys~\cite{cayley1995beaufort, vicsurvey}. A cross-sectional view of the model including SUPL is shown in Fig.~\ref{fig:supl-cross-section}. The muon flux simulated using \textsc{pumas} is shown in Fig.~\ref{fig:pumas_flux} on the left.

After obtaining the simulated acceptance and flux as a function of the angle, the acceptance map in Fig.~\ref{fig:geo_acc} is used to filter the flux distribution, multiplying each flux bin by the corresponding acceptance. The result is shown on the right in Fig.~\ref{fig:pumas_flux}.

\subsection{Geometric acceptance systematic uncertainty}
\label{subsec:geometric_acceptance_sys}
The measurement of $\alpha$ in Eq.~\ref{eq:acceptance} is sensitive to several systematic effects. First, uncertainties in the densities of the materials in the muon transport simulation must be considered, which alter the value of $f_{\rm sim}(\eta, \phi)$. Second, the Landau selection criterion used to measure the raw flux affects the angular acceptance of the telescope for all muons that traverse only part of the full scintillator thickness. This effect would alter $\alpha_{\rm sim}(\eta, \phi)$. Finally, the value of the integral in Eq.~\ref{eq:acceptance} is affected by the orientation of the telescope since the muon flux is not invariant for rotations of the angle $\phi$. The following subsections address the estimation of these uncertainties.

\subsubsection{Edge effect uncertainty}
\label{subsubsec:edge_effect_unc}
When selecting muon events using the Landau fit criterion, a high rejection is achieved for muons that only partially cross the panel thickness. Ideally, this corresponds to deriving the nominal acceptance map using the most stringent geometrical requirement on the muon trajectories. For this reason, the nominal map shown in Fig.~\ref{fig:geo_acc} is obtained for events of type C (see Fig.~\ref{fig:event_vis}). However, this geometrical requirement is not expected to exactly match the Landau criterion, which is based on energy deposition instead. Indeed, the nominal event selection might be more inclusive, {\it e.g.} for those partially contained muons that still lead to considerable energy deposits due to long longitudinal traversals.
The Landau selection has not been further calibrated against muon trajectories of known inclination. Therefore, it might impose a constraint on type C muons themselves. To estimate a symmetrical uncertainty on $\alpha$, we use the difference between the two $\alpha$ values obtained with the nominal $\alpha_{\rm sim}(\eta, \phi)$ map and a modified version where events of type A, B, and D are included. We then take half of this difference to symmetrise the uncertainty, as shown in Table~\ref{tab:acc_sys_tab}.

\begin{figure}[htb]
    \centering
    
    \includegraphics[width=0.49\textwidth]{Acceptances_AllPanels_PostBoundary_PanelsAcceptance.png}
    \includegraphics[width=0.49\textwidth]{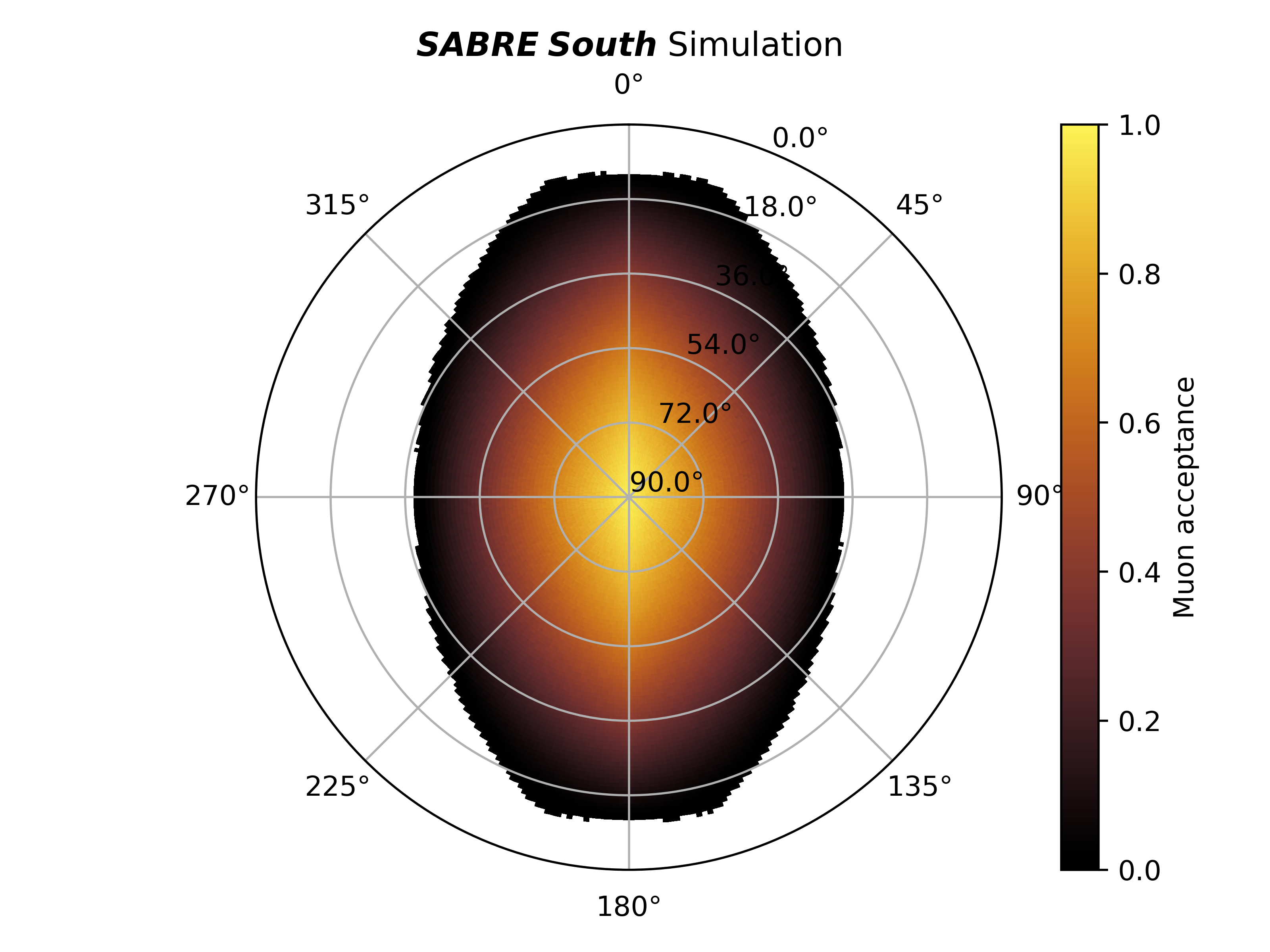}
    
    \caption{Comparison between the nominal (left) and systematically modified (right) telescope geometrical acceptances obtained using simulated events to quantify the edge effect uncertainty. In the nominal map, events are constrained to traverse the full thickness of the top and bottom scintillators, while they are not in the systematically modified case.}

    \label{fig:acc_sys}
\end{figure}

\subsubsection{Material density uncertainty}
\label{subsubsec:material_density_unc}
\begin{figure}[htb]
    \centering
    \includegraphics[width=0.4\textwidth]{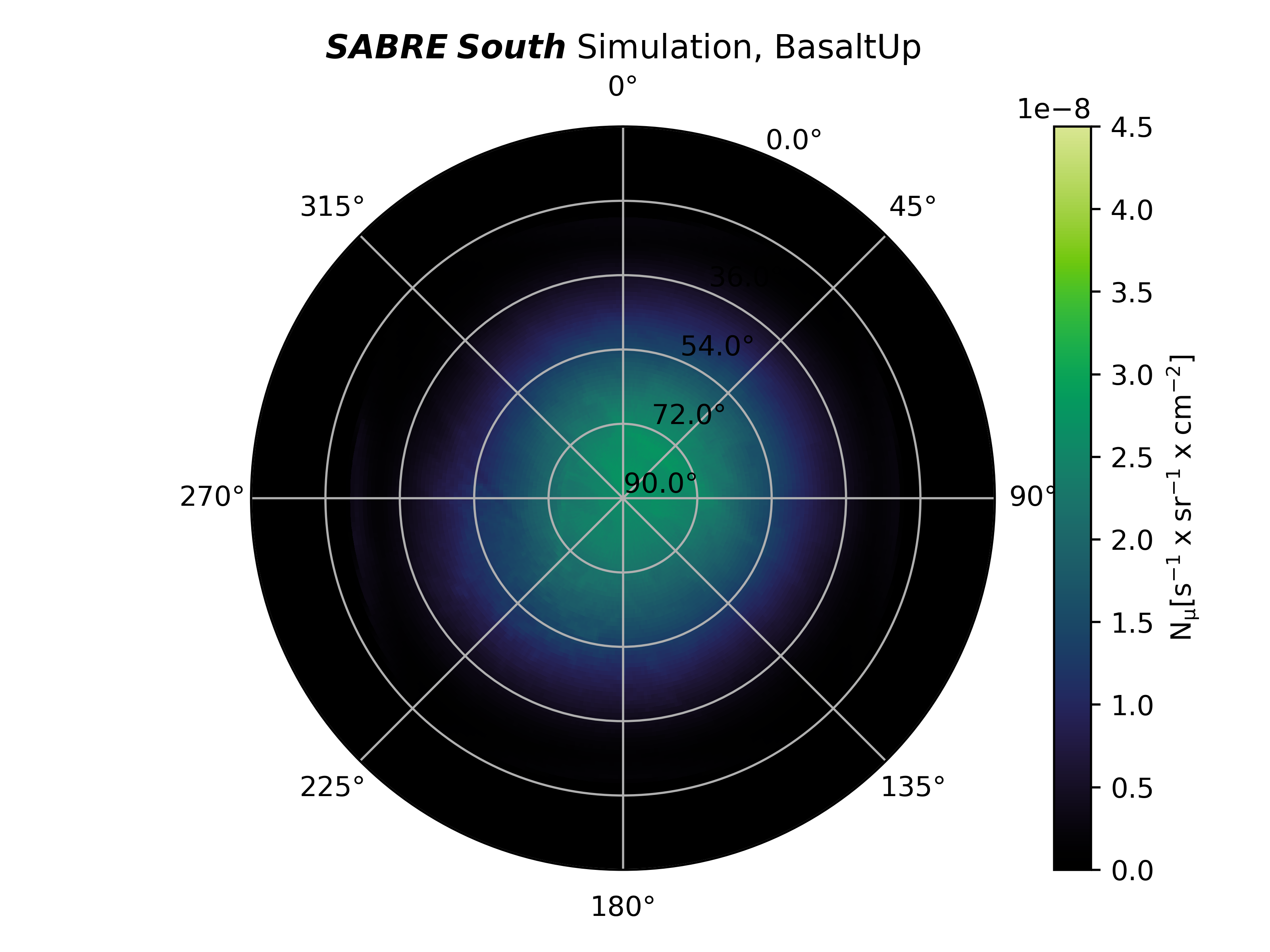}
    \includegraphics[width=0.4\textwidth]{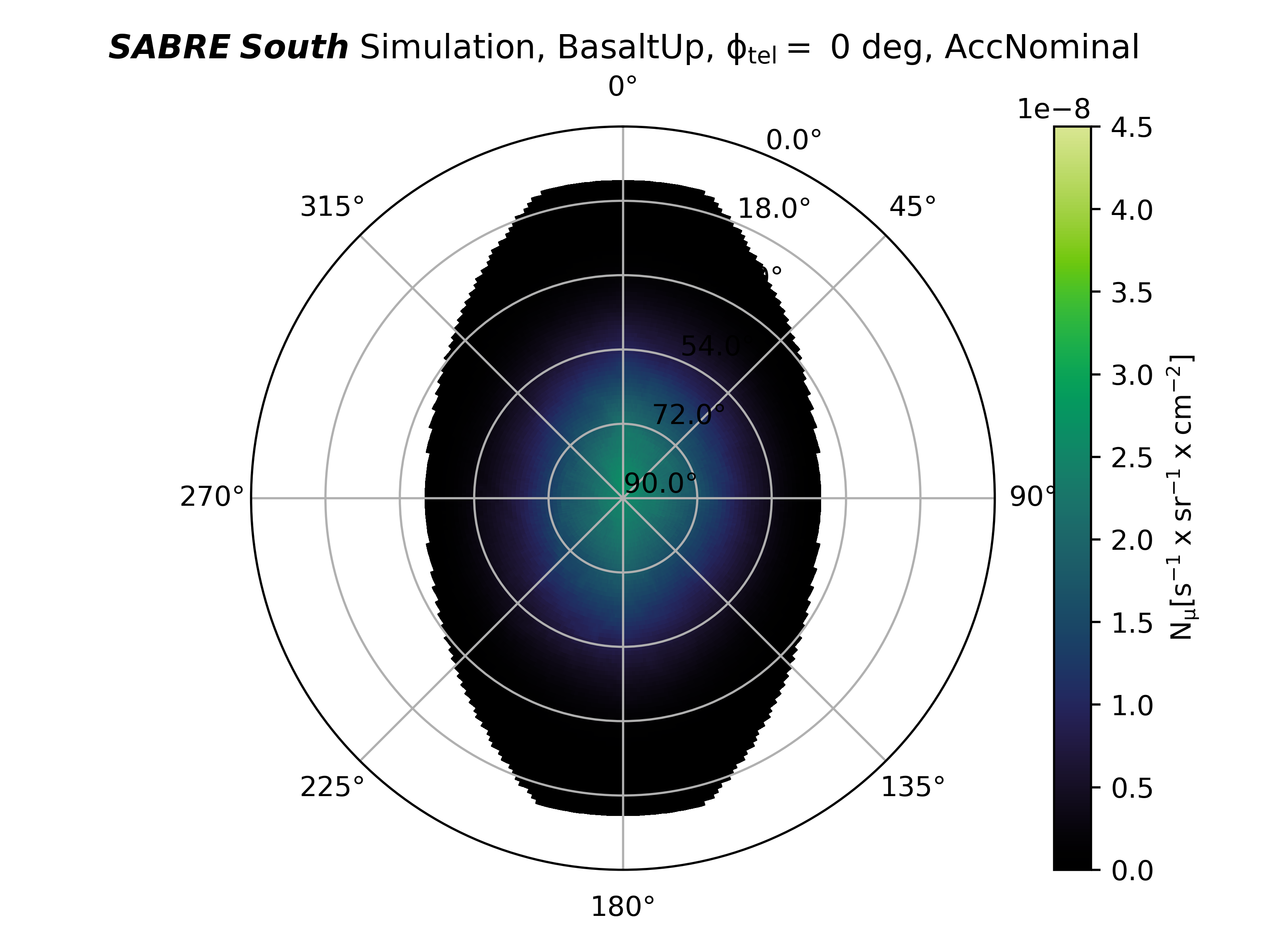}
    
    \includegraphics[width=0.4\textwidth]{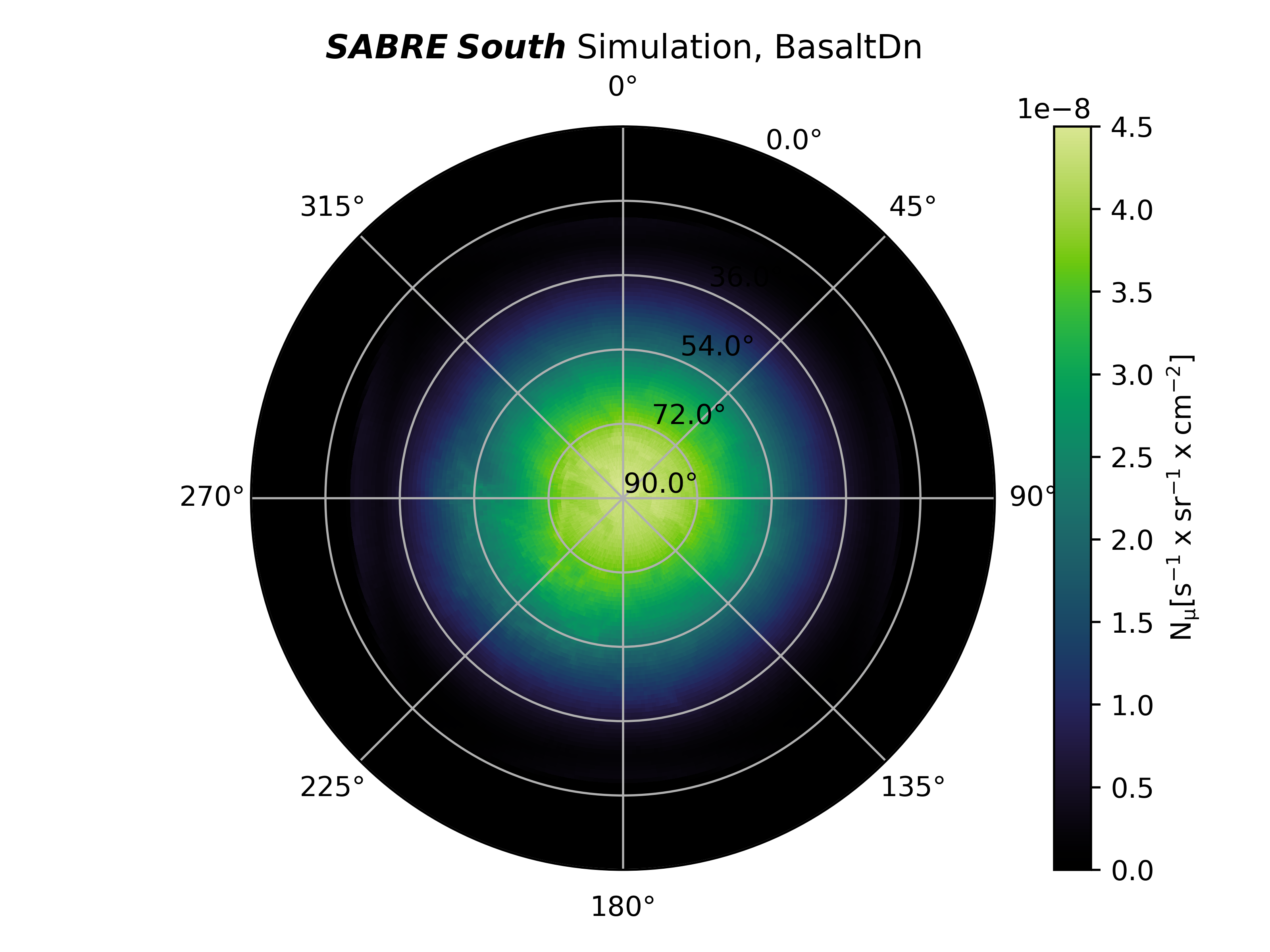}
    \includegraphics[width=0.4\textwidth]{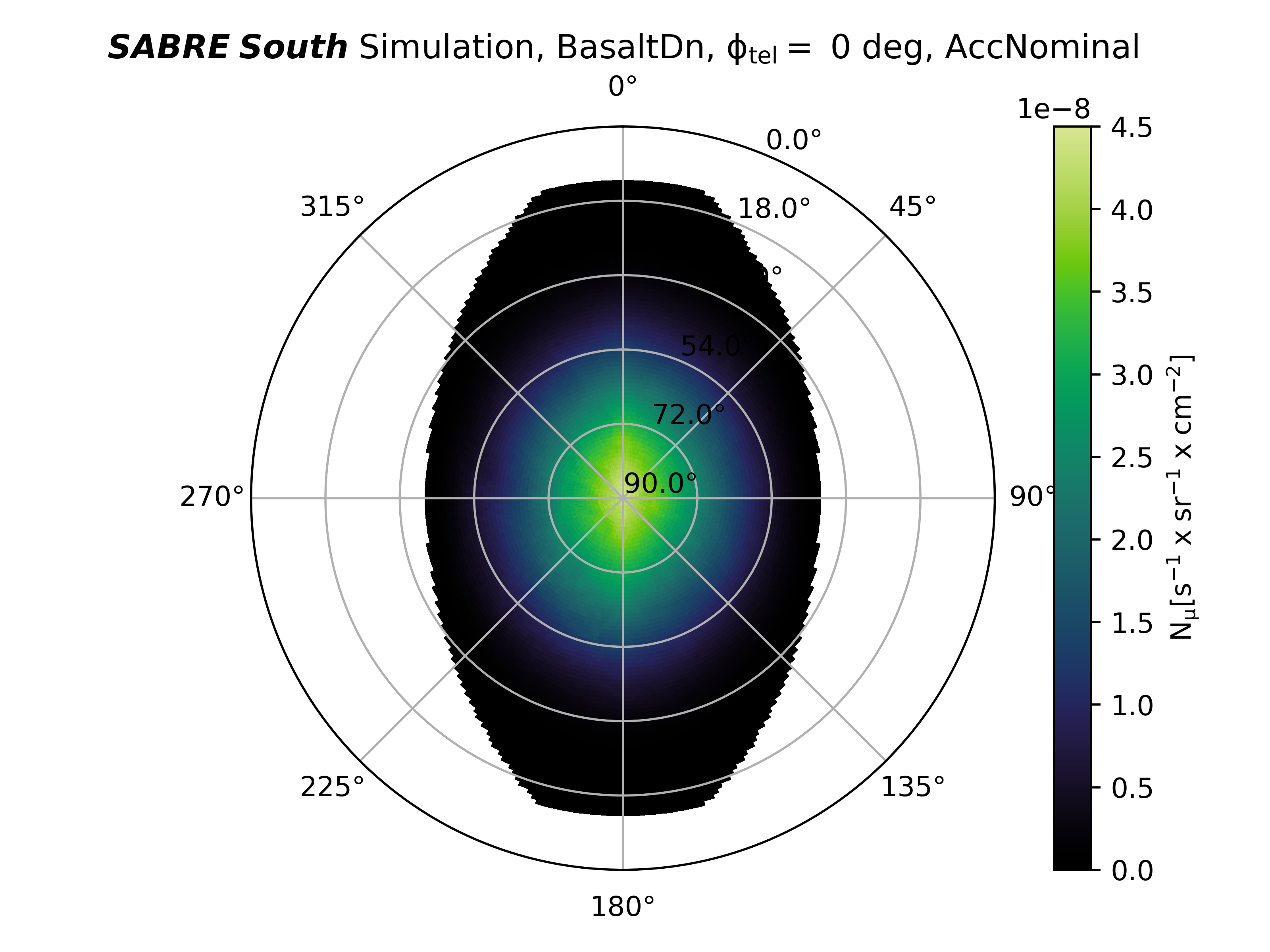}
    
    \includegraphics[width=0.4\textwidth]{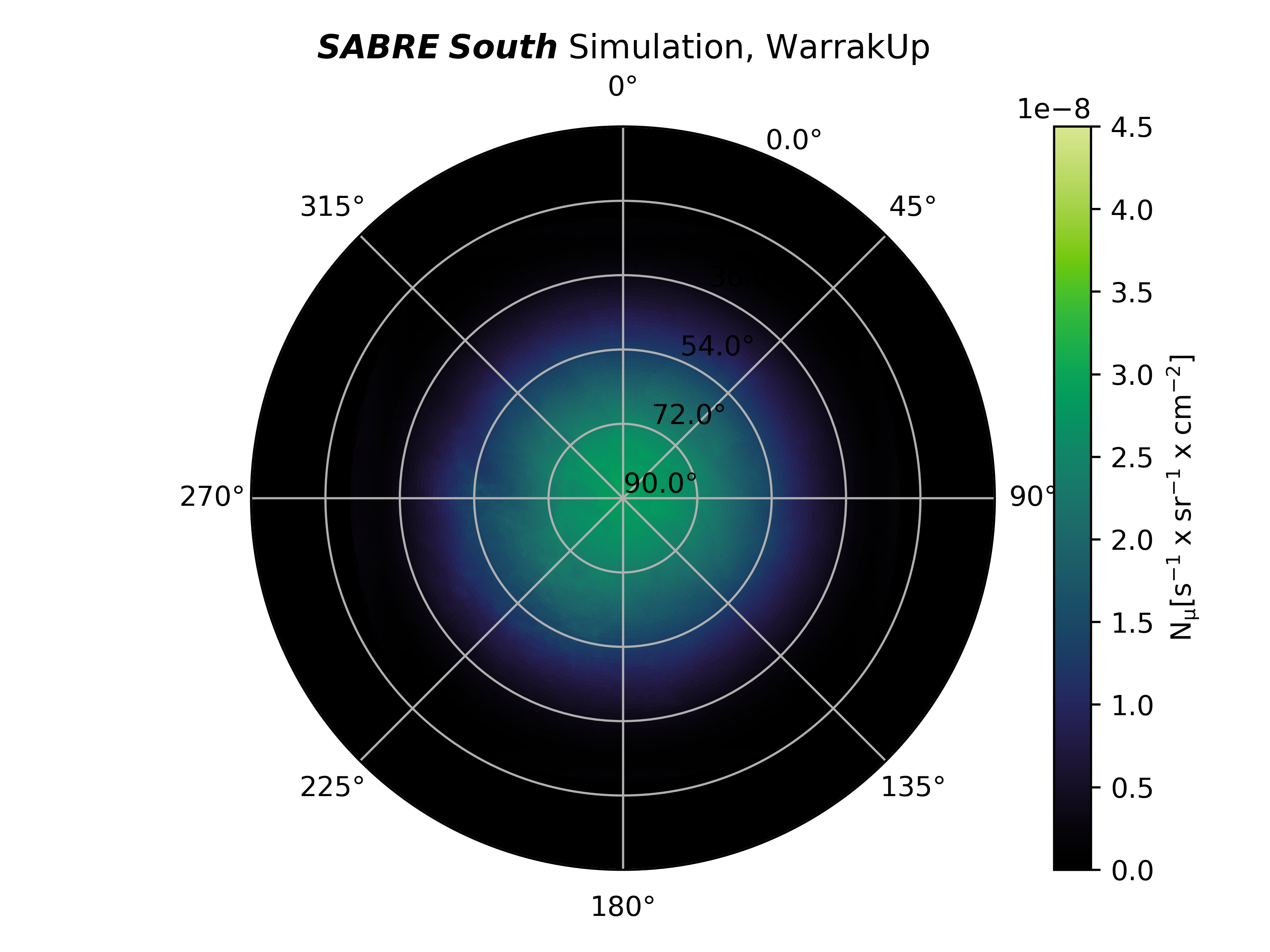}
    \includegraphics[width=0.4\textwidth]{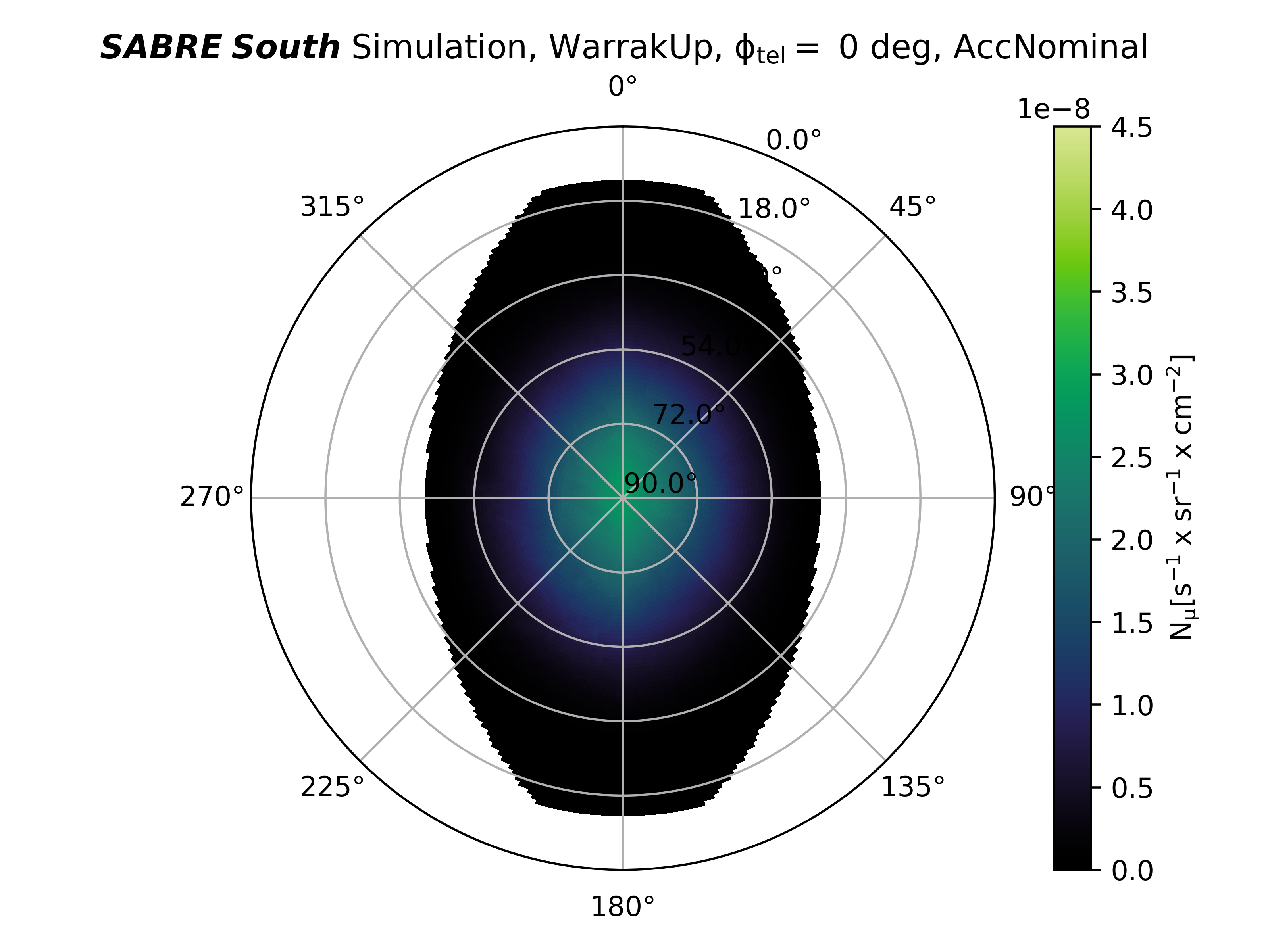}
    
    \includegraphics[width=0.4\textwidth]{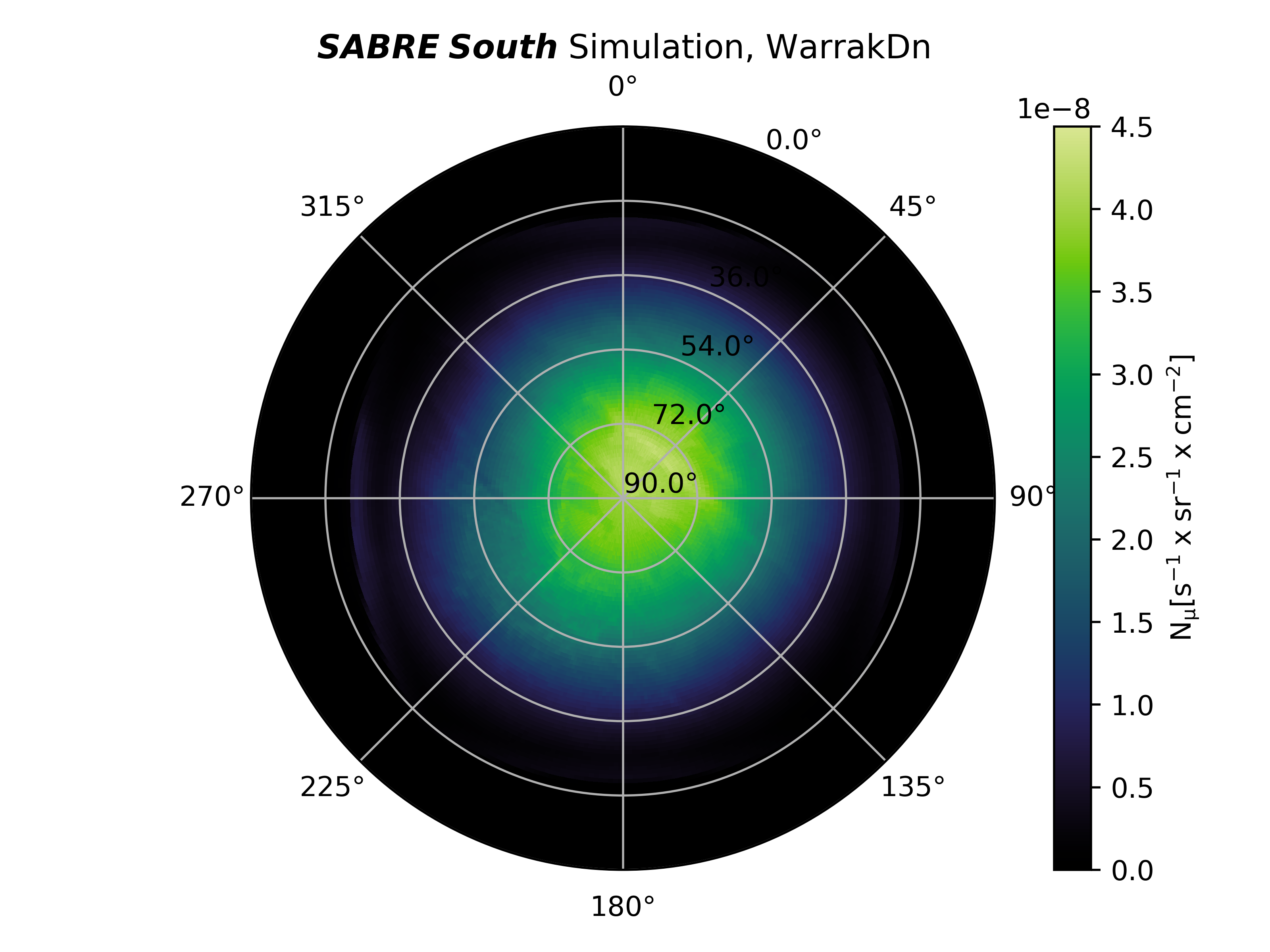}
    \includegraphics[width=0.4\textwidth]{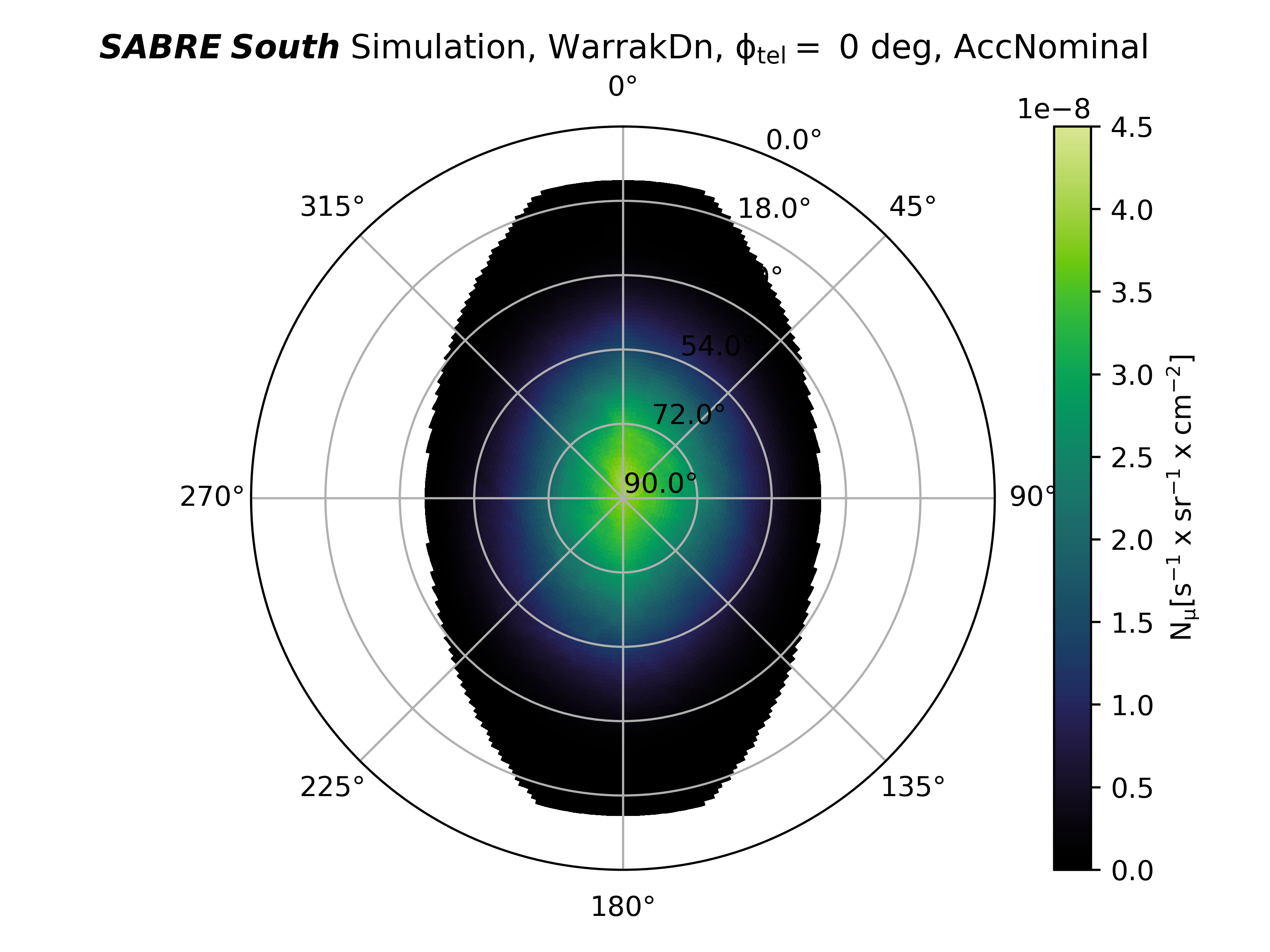}
    
    \caption{Simulated muon flux at SUPL where the density of the basalt and sandstone (Warrak) in the model has been varied up and down by 1$\sigma$. The left-hand plots show a simulation where the geometrical acceptance is 100$\%$, while, in the right-hand plots, the flux is filtered with the nominal acceptance. The nominal orientation of the telescope is assumed, with its length ranging from North to South.}

    \label{fig:pumas_flux_density_sys}
\end{figure}

Uncertainties in the densities of basalt and sandstone in the model used to obtain $f_{\rm sim}(\eta, \phi)$ have a significant impact on the magnitude of the flux. However, $\alpha$ is not significantly affected by changes in flux magnitude, due to the renormalisation factor in the denominator of Eq.~\ref{eq:acceptance}. These two materials have distributions that vary with the two angular coordinates in different ways, and are expected to affect $\alpha$ nonetheless.  
To propagate the effect of density variations on $\alpha$, the two densities are varied up and down within their $1\sigma$ uncertainties, and the flux maps are re-simulated for each variation (Fig.~\ref{fig:pumas_flux_density_sys}). The overall effect on $\alpha$ has been conservatively assumed to be the uncorrelated quadratic summation of each effect. Results are shown in Table~\ref{tab:acc_sys_tab}.   

\subsubsection{Telescope orientation uncertainty}
\label{subsubsec:telescope_orientation_unc}
\begin{figure}[htb]
    \centering
    
    \includegraphics[width=0.4\textwidth]{PlotJob_El1_Az1_Nominal_MuonSimRate_FilteredAz0_AccNominal_Nominal.png}
    \includegraphics[width=0.4\textwidth]{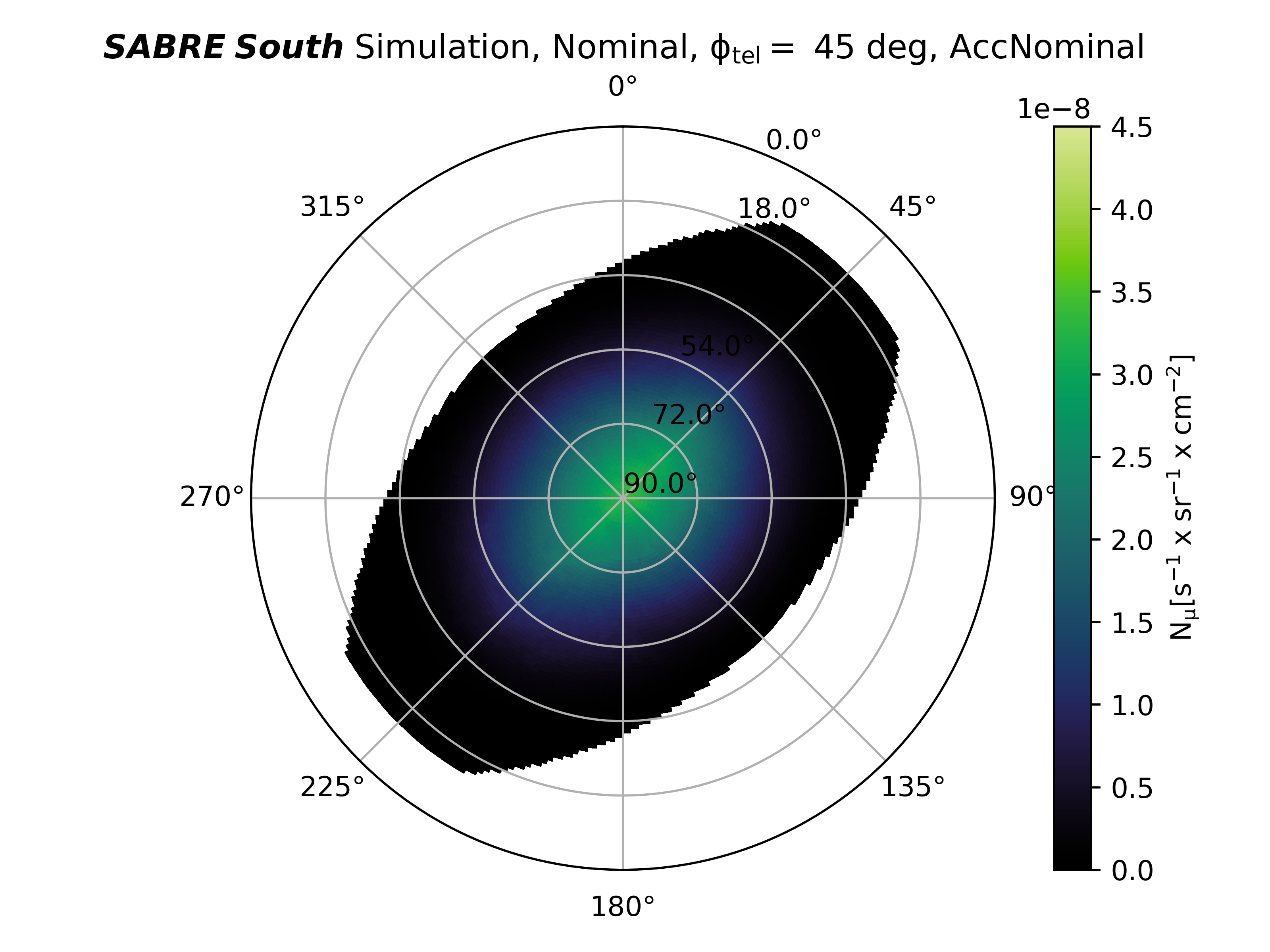}
    \includegraphics[width=0.4\textwidth]{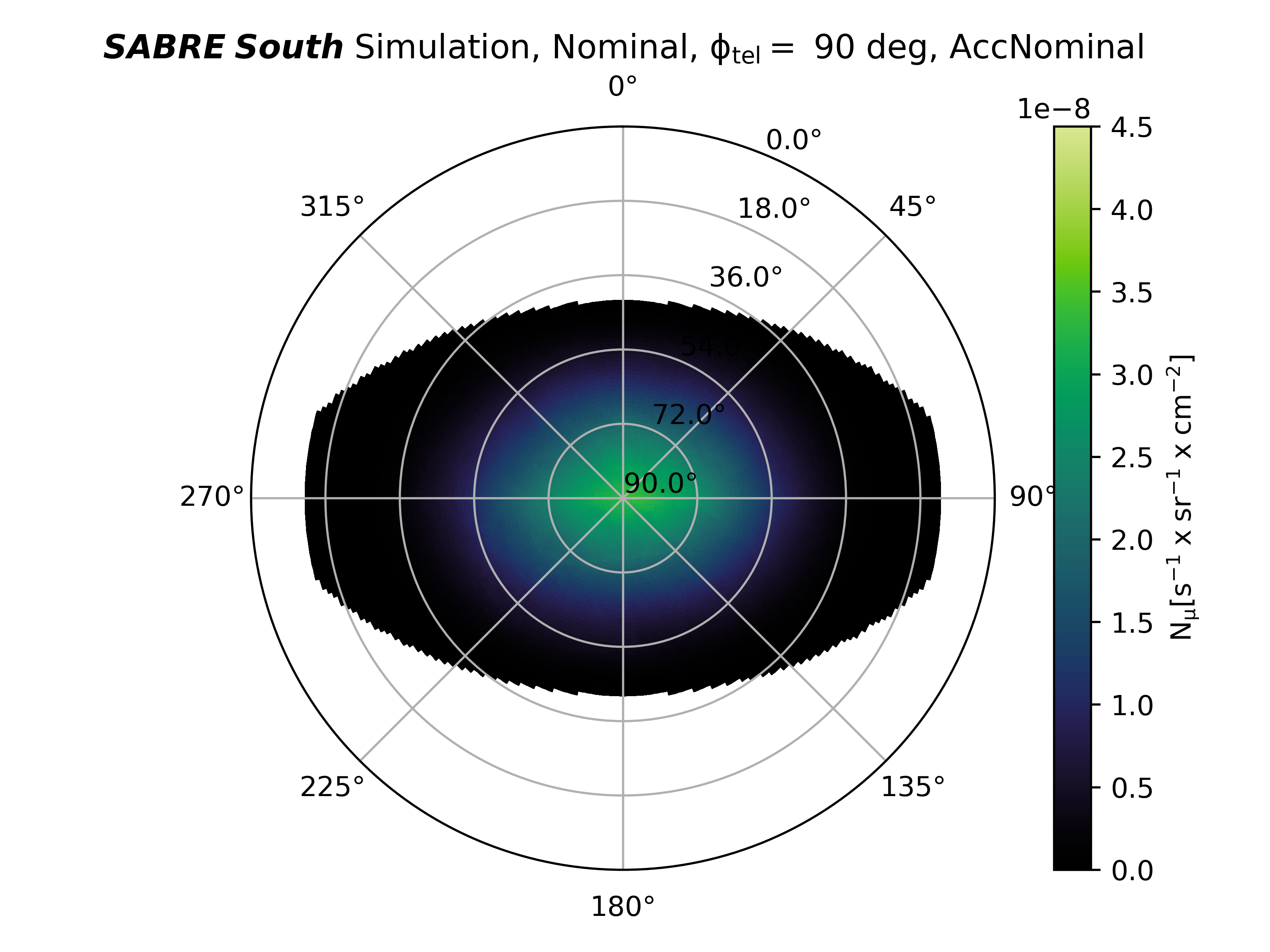}
    \includegraphics[width=0.4\textwidth]{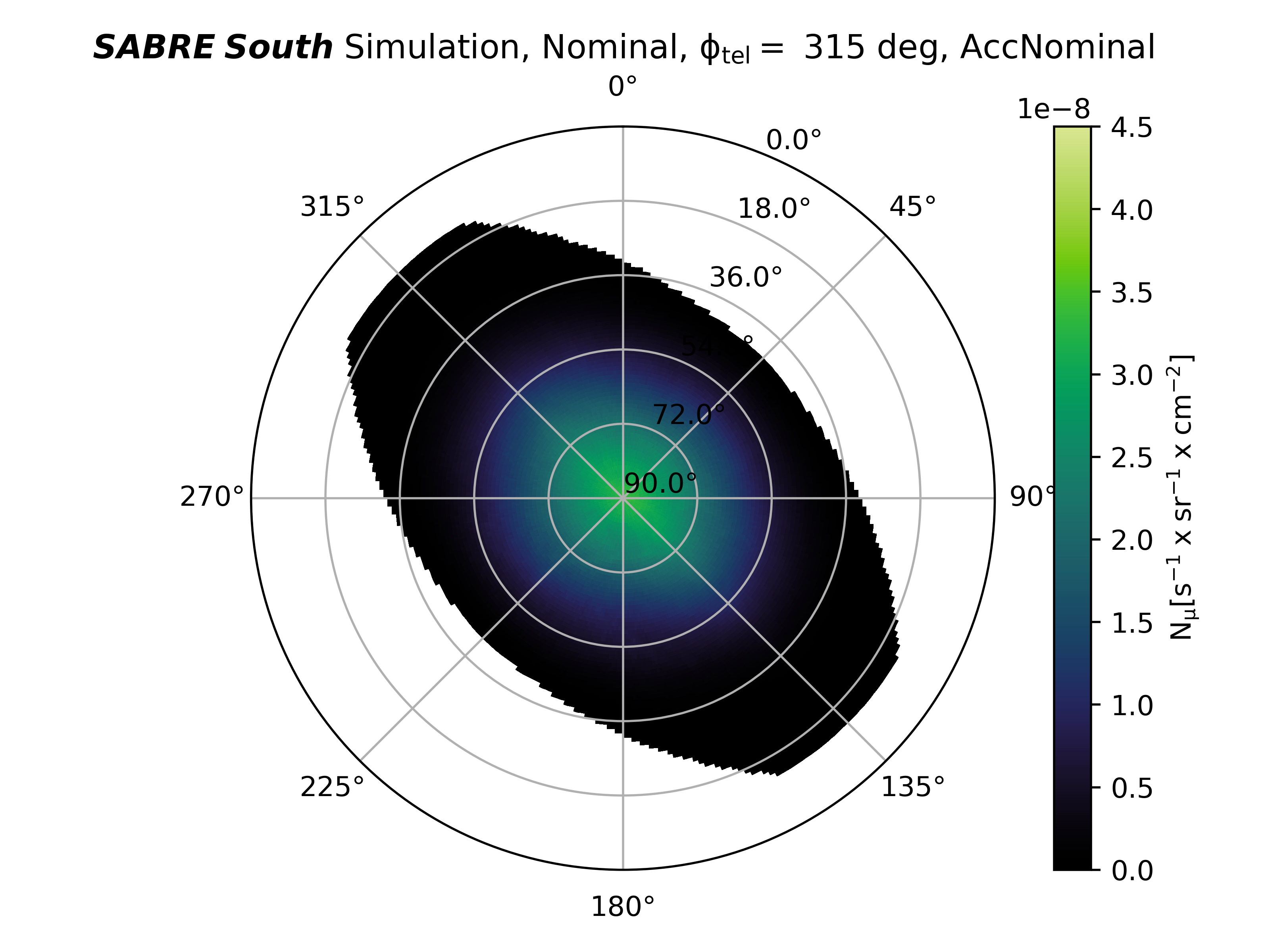}
    
    \caption{Simulated muon flux at SUPL where the orientation of the telescope is rotated. The nominal flux is shown in the upper-left corner, while additional plots consider positive rotations of $\phi=\ang{45},\,\ang{90},\,\ang{315}$.}

    \label{fig:pumas_flux_orientation_sys}
\end{figure}

The telescope orientation at SUPL is affected by a significant uncertainty. Since angular anisotropies are expected in the flux (as discussed in the previous section), different telescope orientations have been considered in deriving $\alpha$. The nominal $\alpha_{\rm sim}(\eta, \phi)$ map is rotated (Fig.~\ref{fig:pumas_flux_orientation_sys}), and the evaluation of $\alpha$ is repeated for orientations at  $\phi=\ang{45},\ang{90},\ang{315}$. The uncertainty is assumed to be the symmetrised maximum variation on $\alpha$, as illustrated in Table~\ref{tab:acc_sys_tab}.

\begin{table}[htb]
\centering
\caption{Summary of systematic uncertainties on $\alpha$ measured using simulations. The table provides a breakdown of individual effects, where each source of uncertainty is investigated through multiple parameter variations. The uncertainty on the edge effect has been symmetrised using its maximal variation. Uncertainties associated with varying material densities are added in quadrature. The uncertainty associated with the telescope orientation is the maximum variation among all systematically rotated orientations.}
\label{tab:acc_sys_tab}
\begin{tabular}{lccc}
\toprule
Systematic effect on $\alpha$          &&& Total systematic uncertainty \\ \hline
Edge effect                   && $+$ 0.032  &      $\pm$ 0.016                       \\ \hline 
\multirow{4}{*}{Density}     &$\rho_{\rm basalt} = 2.81+0.25\,{\rm g/cm^3}$& + 0.002  & \multirow{4}{*}{$\pm$ 0.020}          \\
                             &$\rho_{\rm basalt} = 2.81-0.25\,{\rm g/cm^3}$& + 0.004  &                              \\
                             &$\rho_{\rm Warrak} = 2.74+0.3\,{\rm g/cm^3}$& + 0.016   &                              \\
                             &$\rho_{\rm Warrak} = 2.74-0.3\,{\rm g/cm^3}$& $-$ 0.011   &                              \\ \hline
\multirow{3}{*}{Orientation} &$\phi_{\rm tel}=\ang{45}$& + 0.005   & \multirow{3}{*}{$\pm$ 0.005}            \\
                             &$\phi_{\rm tel}=\ang{90}$& + 0.003   &                              \\
                             &$\phi_{\rm tel}=\ang{315}$& $-$ 0.003  &                              \\ \hline
Nominal  $\alpha$ value                    && 0.483 &               $\pm$ 0.026               \\ 
\bottomrule
\end{tabular}
\end{table}

\section{Measurement of the muon flux}
\label{sec:flux_measurement}
After determining all quantities on the right-hand side of Eq.~\ref{eq:muon_flux_final}, the final muon flux can be calculated. The final result is shown in Table~\ref{tab:muon_flux_summary}. In the table, the muon detection efficiency $\varepsilon$ is measured as the product of two single-panel efficiencies. Additionally, before applying the efficiency and acceptance corrections, the raw fluxes from the two telescopes are combined using the following weighted average:
\begin{equation}
\label{eq:muon_flux_weighted_average}
    f^{\rm raw} = \frac{f^{\rm raw}_{1} / \sigma^2_1 + f^{\rm raw}_{2} / \sigma^2_2}{1/\sigma^2_1 + 1/\sigma^2_2},
\end{equation}
where $\sigma_{1,2}$ are statistical uncertainties and the variances are used to define the weights in the average.

\begin{table}[htp]
\centering
\caption{Summary of the muon flux measurement at SUPL. All quantities relevant to the final measurement are shown, including their statistical and systematic uncertainties. The fluxes for Telescope 1 and 2 are combined using a weighted average.}
\label{tab:muon_flux_summary}
\begin{tabular}{lccc}
\toprule
  & \multirow{2}{*}{Nominal value} & \multicolumn{2}{c}{Uncertainty} \\
  &                                & statistical     & systematic    \\ \hline
Telescope 1 $f^{\rm raw}\,[{\rm s^{-1} cm^{-2}}]$ &      $3.03\times10^{-8}$  &  $\pm\,0.02\times10^{-8}$ &  \\
Telescope 2 $f^{\rm raw}\,[{\rm s^{-1} cm^{-2}}]$ &      $3.02\times10^{-8}$  &  $\pm\,0.02\times10^{-8}$ &  \\
Average $f^{\rm raw}\,[{\rm s^{-1} cm^{-2}}]$     &      $3.03\times10^{-8}$  &  $\pm\,0.02\times10^{-8}$ &  \\
$\varepsilon$             &      0.989              &  - & $\pm$ 0.003 \\
$\alpha$                  &      0.483               &   -             & $\pm$ 0.026 \\ \hline
$f\,[{\rm s^{-1} cm^{-2}}]$                       &      $6.33\times10^{-8}$ &  $\pm\,0.04\times10^{-8}$ & $\pm\,0.35\times10^{-8}$ \\
\bottomrule
\end{tabular}
\end{table}

\begin{figure}
    \centering
    \includegraphics[width=0.8\linewidth]{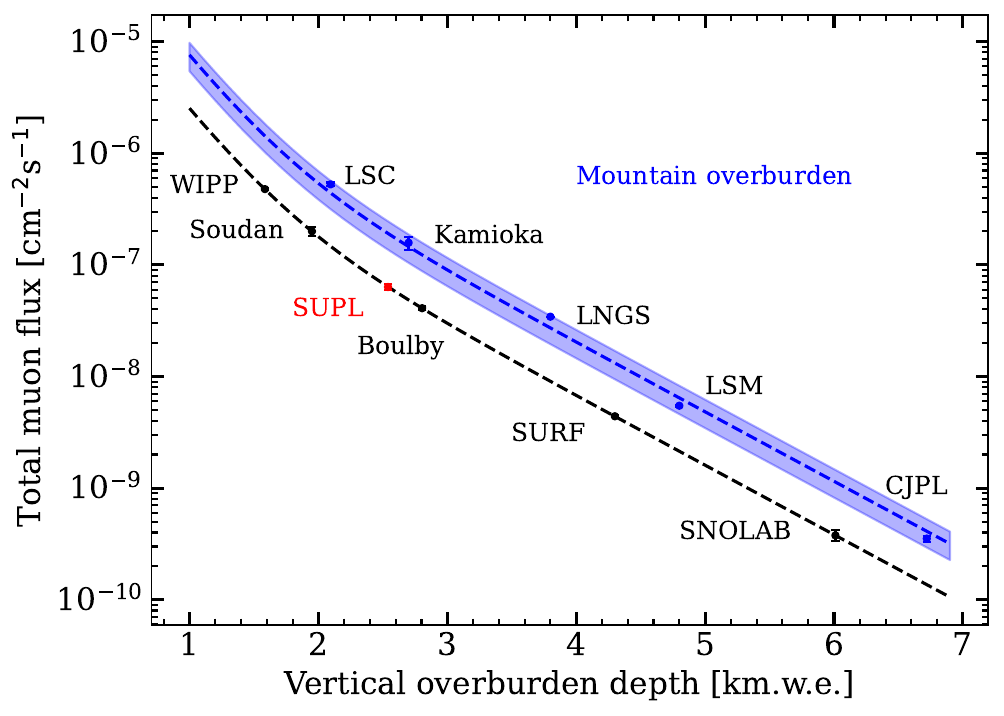}
    \caption{The total muon flux as a function of vertical overburden depth. The black dashed line is from Mei's model \cite{Mei:2005gm}. The blue dashed line is obtained by scaling Mei's model to adapt it for cases under mountain overburdens~\cite{JNE:2020bwn}. The blue band shows the uncertainty of the scaling. Data points are measurements from WIPP~\cite{Esch:2004zj}, LSC~\cite{Trzaska:2019kuk,Morales:2004ae}, Soudan~\cite{Zhang:2014jsq}, Kamioka~\cite{KamLAND:2009zwo}, Boulby~\cite{Reichhart:2013xkd}, LNGS~\cite{Borexino:2018pev}, SURF~\cite{MAJORANA:2016ifg}, LSM~\cite{EDELWEISS:2013kzp}, SNOLAB~\cite{SNO:2009oor}, and CJPL~\cite{JNE:2020bwn}. }
    \label{fig:muonflux_depth}
\end{figure}

The final muon flux measurement is shown with other results in the literature in Fig.~\ref{fig:muonflux_depth}. The blue points represent underground laboratories located under a mountain overburden, and the black points represent those located under a flat overburden.
In the plot, curves are obtained using the so-called $slant\;depth$ parameterisation~\cite{Mei:2005gm, JNE:2020bwn}, while data points from the various collaborations take into account the complex geometries of their respective overburden.

The expected flux value calculated from simulated events is
\begin{equation}
\label{eq:muon_flux_simulated}
    f_{\rm sim} = (4.8\,\pm\,2.3)\times10^{-8}\,[{\rm s^{-1} cm^{-2}}],
\end{equation}

which is in good agreement with the measured value 

\begin{equation}
\label{eq:muon_flux_meas}
    f = (6.33\,\pm\,0.04_{\rm stat}\,\pm\,0.35_{\rm sys})\times10^{-8}\,[{\rm s^{-1} cm^{-2}}].
\end{equation}

The high uncertainty on the expected flux is dominated by the systematic uncertainty on the density of the material of the overburden. Note that this effect is attenuated in the measured value due to the form of Eq.~\ref{eq:acceptance} where the simulated flux appears in the numerator and denominator. 

\section{Conclusions}
The muon flux at SUPL has been measured using two telescopes constructed from eight scintillator panels of the SABRE South experiment muon veto system. The measurement uses approximately 236 days of data collected in 2024 and 2025. The final result, corrected for experimental effects including muon detection efficiency and telescope acceptance, is:  
    $f = (6.33\,\pm\,0.04_{\rm stat}\,\pm\,0.35_{\rm sys})\times10^{-8}\,[{\rm s^{-1} cm^{-2}}]$.

The muon flux is known to be affected by the atmospheric temperature and pressure. This result represents the average muon flux over the data-taking period. The expected muon modulation amplitude at SUPL is estimated to be approximately $0.8\%$, which is small compared to the systematic uncertainty.

\section*{Acknowledgements}
\begin{sloppypar}
Work at Stawell Underground Physics Laboratory is supported by the SUPL Board. The SABRE South program is supported by the Australian Government through the Australian Research Council (Grants: CE200100008, LE250100018, LE190100196, LE170100162, LE160100080, DP190103123, DP170101675, LP150100705), the Laby Foundation at the University of Melbourne, and the University of Melbourne Faculty of Science, with additional support from Australian Government Research Training Program Scholarships. This research used resources provided by The University of Melbourne’s Research Computing Services and Petascale Campus Initiative. The authors would like to thank Stawell Gold Mines for providing support throughout operations and for providing the overburden data for muon flux simulations. 
\end{sloppypar}

\bibliographystyle{elsarticle-num} 
\bibliography{sabre}

\end{document}